\documentclass[12pt,fleqn]{article}
\topmargin - 0.8cm

\catcode`@=11
\@addtoreset{equation}{section}
\catcode`@=12
\mathchardef\SGamma="7100

\begin{document}

\title{\bf Effective equations in quantum cosmology}
\author{~A.~O.~Barvinsky $^a$\thanks{Email address: barvin@td.lpi.ac.ru}
 ~and  ~D.~V.~Nesterov $^b$\thanks{Email address: nesterov@td.lpi.ru}}
\date{}
\maketitle
\begin{center}
\hspace{-8mm}$^a${\em
Theory Department, Lebedev Physics Institute and Lebedev Research Center in
Physics, Leninsky Prospect 53,
Moscow 117924, Russia}\\
\hspace{-8mm}$^b${\em
Moscow State University, Physics Faculty,
Department of Theoretical Physics \\
Moscow 119899, Russia.}
\end{center}

\begin{abstract}
We develop a general framework for effective equations of expectation values
in quantum cosmology and pose for them the quantum Cauchy
problem with no-boundary and tunneling wavefunctions. Cosmological
configuration space is decomposed into two sectors that give
qualitatively different contributions to the radiation currents in
effective equations. The field-theoretical sector of inhomogeneous modes
is treated by the method of Euclidean effective action, while the
quantum mechanical sector of the spatially homogeneous inflaton is
handled by the technique of manifest quantum reduction to gauge
invariant cosmological perturbations. We apply this framework in the
model with a big negative non-minimal coupling, which incorporates a
recently proposed low energy (GUT scale) mechanism of the quantum origin
of the inflationary Universe and study the effects of the quantum inflaton
mode.
\end{abstract}

\section{Introduction}
\hspace{\parindent}
This paper is a sequel to the previous work \cite{efeq} on quantum
dynamics of the early Universe starting from initial conditions inspired by
quantum cosmology. These initial conditions, encoded in the no-boundary
\cite{HH,H,VilNB} and tunneling \cite{tun} quantum states, can be a source
of the inflationary scenario at the low (typically GUT) energy scale
\cite{qsi,qcr}, compatible with the observational status of inflation theory.
In \cite{qsi,qcr} the initial conditions were found as a sharp
peak in the probability distribution of the quantum scalar field whose
expectation value simulates the effective Hubble constant and drives
inflation. The parameters of this peak -- its location and quantum width --
are suppressed relative to the Planckian values by a small factor coinciding
with the magnitude of the CMBR anisotropy, $\Delta T/T\sim 10^{-5}$,
according to normalization on COBE \cite{COBE,Relict}. In \cite{noanthr,open}
these results were extended to the open model based on the Hawking-Turok
instanton \cite{HawkTur}.

Although, apriori these features make this model attractive, serious
objections usually arise regarding the present day status of initial
conditions in cosmology. Their issue in inflation theory is a subject of
strong debate in current literature \cite{Guth} and numerous meetings on
structure formation of the Universe. The possibility of quantum cosmological
imprint on the observational data was called in question either from the
viewpoint of the self-reproducing inflation scenario \cite{inflation,Lindop}
or from the viewpoints of the eternal inflation \cite{Guth} and anthropic
principle \cite{Vilanthr}. However, these objections become legitimate
only when the conditions of self-reproduction and eternal inflation or the
dominance of anthropic probabilities are guaranteed. In their turn, the
realization of these conditions depends on cosmological state. Typically,
they require the initial Planckian energy scale, a very long duration of
the inflationary epoch, etc., and as a rule go beyond the scope of reliable
perturbative methods. The model of \cite{efeq,qsi,qcr} avoids all these
aspects, because it represents maybe the first example of the {\em low-energy
phenomenon of the quantum birth of the Universe}. Thus, it is likely to be a
reasonable candidate for the cosmological quantum state deserving further
studies.

The dynamical consequences of the probability peak in the
distribution of the inflaton were first studied at the level of classical
equations \cite{qsi,qcr}. Then they were reexamined at the level of
effective equations for quantum expectation values of fields \cite{efeq}.
However, the analyses of \cite{efeq} was not complete -- the quantum
contribution to effective equations included all the modes {\em except} the
main spatially homogeneous mode related to quantum fluctuations of the
inflaton field itself. This is the main spatially global degree of freedom
in cosmology. It is responsible for the quantum fluctuations of the
homogeneous (minisuperspace) background on top of which all the other
inhomogeneous modes dynamically evolve.

As was understood in \cite{efeq}, the contribution of this mode
requires the calculational technique different from the technique for
inhomogeneous modes. The latter
is strongly facilitated by the method of {\em Euclidean effective action}
which is based on specific properties of their quantum state on the
quasi-DeSitter background -- Euclidean DeSitter invariant vacuum
\cite{BAllen}. On the contrary, the quantum state of the inflaton is
irrelevant to this vacuum and its contribution cannot be obtained by the
analytic continuation from the Euclidean effective action. Rather, this
state is determined by the peak of the inflaton probability distribution
and can be approximated by the gaussian packet. Direct quantum averaging
with respect to this packet is required for obtaining the contribution of
the inflaton mode to effective equations. This step, however, should be
preceded by several other calculational steps involving the Hamiltonian
reduction to the physical sector, setting the quantum Cauchy problem in
the minisuperspace sector of the system, etc. Thus, the goal of the present
paper is to pose the quantum Cauchy problem in cosmology for the
no-boundary and tunneling quantum states, derive the one-loop contribution
of the homogeneous mode to effective equations and analyze its influence
on the inflationary dynamics.

One of particular questions, to be clarified in this work, is to what an
extent this contribution can change the conclusions of our previous papers
\cite{efeq,qcr}. In the spatially closed model with the no-boundary and
tunneling cosmological states these conclusions were pretty stringent. The
no-boundary quantum state can give rise only to the eternally long inflation,
while the finite inflation stage with the conventional exit
to the matter-dominated Universe exists only for the tunneling
state\footnote{In open cosmology based on the Hawking-Turok
instanton \cite{HawkTur} the initial conditions for finite inflation
stage can be realized for the no-boundary state \cite{noanthr,open}, but the
models of \cite{open,HawkTur} with all their merits and disadvantages
go beyond the scope of this paper.}. These conclusions give an
undoubtful preference to the tunneling state from
the viewpoint of the observational cosmology, but, at the fundamental level,
the tunneling state has an intrinsic problem if one goes beyond the tree-level
approximation\footnote{Point is that the classical Euclidean action enters the
exponential of the tunneling wavefunction with the ``wrong'' sign -- opposite
to that of the loop corrections \cite{tvsnb}. This mismatch invalidates the
conventional renormalization procedure of absorbing the ultraviolet
divergences into the redefinition of classical coupling constants
\cite{Rubakov}. This inconsistency does not break the results of
\cite{efeq,qsi,qcr} (heavily relying on loop effects) for accidental
reasons -- renormalization ambiguous part of the distribution function in
the slow roll approximation factors out as an inert overall normalization.
However, it shows up beyond this approximation and, thus, persists at the
fundamental level.}. So, there is a hope that the inclusion of quantum
fluctuations in the minisuperspace sector can render the no-boundary state
viable from the
viewpoint of the inflationary cosmology. Unfortunately, as we shall see, this hope
does not realize -- the effect of this mode turns out to be too small.
This result, however, is model dependent, and the application of the general
framework of this paper might lead to other conclusions in alternative models
with well-defined quantum states of the inflaton.

It should be emphasized, that despite intensive studies on the inflationary
dynamics, the quantum nature of the homogeneous inflaton mode has not yet
been completely understood. This mode is peculiar because of its ghost
nature -- its kinetic term enters the action with the wrong sign. This fact
was emphasized in \cite{GMTS}, but its cosmological consequences have not yet
properly been examined. There is a viewpoint that this mode should at all
be excluded from the quantum perturbations spectrum on the physical grounds
that it is always beyond the horizon and not directly observable \cite{Star}.
This approach can hardly be justified, because this mode is in some sense the
most fundamental one, for it determines the homogeneous background on top of
which all observable perturbations are unraveling. The energy of
this quantum mechanical mode in view of its ghost nature is not positive
definite, and it leads to peculiar back reaction phenomena which we are
going to discuss here. Despite their actual small magnitude in our
model, in other cases they might essentially contribute to the
cosmological evolution. As we shall see, quantum fluctuations of the
homogeneous inflaton have the equation of state $p+\varepsilon=0$ and, thus,
it can even be a candidate for the present day cosmological constant,
instead of quintessence. This is one more motivation for our studies.

The organization of the paper is as follows. In Sect. 2 we briefly
recapitulate the results of \cite{qsi,qcr,tvsnb} for the model with a
big negative non-minimal coupling of the inflaton, $-\xi R\varphi^2/2$,
$-\xi=|\xi|\gg 1$, having a quasi-quartic potential $V(\varphi)=
\lambda\varphi^4/4+...\,$. We formulate the mechanism of one-loop radiative
corrections due to which the inflaton distribution acquires a
sharp peak. This peak gives rise to initial conditions for inflation and
quantum fluctuations contributing to the radiation currents -- quantum
terms in the effective equations. In Sect. 3 we reveal the general
structure of these currents and their gauge invariance properties. In
particular, their decomposition into the contributions of two sectors is
presented: the quantum mechanical sector of the homogeneous inflaton and the
field-theoretical sector of inhomogeneous modes. The Euclidean effective
action method is formulated for

the calculation of the latter, while the former is supposed to be obtained
by direct quantum averaging with respect to the peak-like wavefunction of the
inflaton. In Sect. 4 we go over to the generic model with minimally coupled
inflaton. For this model we pose the Cauchy problem for the classical
solution -- the tree-level approximation for expectation values. We perform
quantum reduction of the wavefunction from minisuperspace (of the scale
factor and inflaton) to the physical subspace, the latter playing the role of
the Cauchy surface in minisuperspace at which quantum initial conditions
are posed. For the no-boundary and tunneling wavefunctions this subspace is
chosen as a domain corresponding to the nucleation of the Lorentzian
quasi-DeSitter spacetime from the Euclidean one. In Sect. 5 a similar
quantum Cauchy problem is posed for cosmological perturbations. Their
physical reduction is briefly presented along the lines of \cite{GMTS}.
Perturbations of minisuperspace variables -- scale factor, lapse and
inflaton -- are expressed in the Newton gauge in terms of the invariant
physical variables of \cite{GMTS}. The latter, in their turn, are found as
operators in the representation of the physical wavefunction on the Cauchy
surface of the above type. In Sect. 6 we get back to the non-minimal model
of Sect. 2 and discuss its reparametrization to the Einstein frame which
allows one to transform its quantum Cauchy problem to that of the
minimal model of Sects. 4 - 5. We also describe here the calculation of
radiation currents and derive the expression for the rolling force in
the effective inflaton equation. Sects. 7 and 8 present the resulting
effects respectively at the beginning of inflation and at late steady
inflationary stage. Concluding section contains a brief summary of
results and prospective implications of its technique.

\section{Quantum origin of the Universe as a low-energy phenomenon}
\hspace{\parindent}
The starting point of our considerations is the assumption that quantum
cosmology can give initial conditions for inflation, which in their turn
determine main cosmological parameters of the observable Universe, including
the density parameter $\Omega$. The requirement of $\Omega>1$ in closed
cosmology gives the bound on the e-folding number $N$ -- the logarithmic
expansion coefficient for the scale factor $a$ during the inflation stage
with a Hubble constant $H=\dot a/a$,
        \begin{eqnarray}
        N=\int_0^{t_F}dt H                     \label{1.1}
        \end{eqnarray}
(with $t=0$ and $t_F$ denoting the beginning and the end of inflation
epoch). This bound reads $N\geq 60$ \cite{HawkTur}. On the other hand,
the value of $N$ is directly related to the initial conditions for inflation
-- initial value of the inflaton, $\varphi_I$,
        \begin{eqnarray}
        N\simeq -\int_0^{\varphi_I}
        d\varphi\,\frac{H(\varphi)}{\dot\varphi}.      \label{1.2}
        \end{eqnarray}
In the chaotic inflation model the effective Hubble constant $H=H(\varphi)$
is generated by the inflaton and, therefore, all the parameters
of the inflationary epoch can be found as functions of $\varphi_I$. This
initial condition belongs to the quantum domain, i.e. it is subject to the
quantum distribution following from the cosmological wavefunction. If this
distribution has a sharp probability peak at certain $\varphi=\varphi_I$,
then this value serves as the initial condition for inflation.

There are two known quantum states that lead in the semiclassical
regime to the closed inflationary Universe -- the no-boundary
\cite{HH,H,VilNB} and tunneling \cite{tun} wavefunctions. They both
describe quantum nucleation of the Lorentzian quasi-DeSitter spacetime
from the Euclidean (positive signature) hemisphere -- the gravitational
instanton responsible for the classically forbidden state of the
gravitational field. In the tree-level approximation they generate the
distribution functions
        \begin{eqnarray}
        \rho^{\rm tree}_{NB,T}(\varphi)\sim
        \exp\big[\mp I(\varphi)\big],             \label{1.3}
        \end{eqnarray}
where $I(\varphi)$ is the Euclidean action of this instanton with
the inflaton $\varphi$. These functions are unnormalizable in the
high-energy domain $\varphi\to\infty$ and generally devoid of the
observationally justified probability peaks. However, by including
quantum loop effects and applying the theory to a particular model
with strong non-minimal curvature coupling of the inflaton one can
qualitatively change the situation --  generate a sharp probability peak at
GUT energy scale satisfying the above bound on $N$ \cite{qsi,qcr,tvsnb}.
The basic formalism underlying this result is as follows.

Beyond the tree level the distribution $\rho_{NB,T}(\varphi)$ is not
just the square of the cosmological wavefunction (\ref{1.3}). It
becomes the diagonal element of the reduced density matrix obtained by
tracing out all degrees of freedom but $\varphi$. As shown in
\cite{tvsnb,norm,tunnel,BarvU} in the one-loop approximation it reads
        \begin{eqnarray}
        \rho_{\rm NB,T}(\varphi)\sim\exp[\mp I(\varphi)-
        \mbox{\boldmath$\SGamma$}_{\rm 1-loop}(\varphi)],     \label{1.4}
        \end{eqnarray}
where the classical action is amended by the Euclidean effective action
$\mbox{\boldmath$\SGamma$}_{\rm 1-loop}(\varphi)$ of all quantum fields
that are integrated out. This action is calculated on the same instanton,
and its contribution
can qualitatively change predictions of the tree-level theory due to the
logarithmic scaling behavior of $\mbox{\boldmath$\SGamma$}(\varphi)$. On
the instanton of the size $1/H(\varphi)$ it looks like
$\mbox{\boldmath$\SGamma$}(\varphi)\sim Z\ln H(\varphi)$, where $Z$ is the
total anomalous scaling of all quantum fields in the model.

The model of \cite{qsi,qcr} has the graviton-inflaton sector with
a big negative constant $-\xi=|\xi|\gg 1$ of nonminimal curvature coupling,
        \begin{equation}
        S[g_{\mu\nu},\varphi]
        =\int d^4x\,g^{1/2}\left(\frac{m_{P}^{2}}{16\pi} R(g_{\mu\nu})
        -\frac{1}{2}\xi\varphi^{2}R(g_{\mu\nu})
        -\frac{1}{2}(\nabla\varphi)^{2}
        -\frac{1}{2}m^{2}\varphi^{2}
        -\frac{\lambda}{4}\varphi^{4}\right),      \label{1.5}
        \end{equation}
and generic GUT sector of Higgs $\chi$, vector gauge $A_\mu$ and spinor
fields $\psi$ coupled to the inflaton via the interaction term
        \begin{eqnarray}
        S_{\rm int}
        =\int d^4x\,g^{1/2}\Big(\sum_{\chi}\frac{\lambda_{\chi}}4
        \chi^2\varphi^2
        +\sum_{A}\frac12 g_{A}^2A_{\mu}^2\varphi^2+
        \sum_{\psi}f_{\psi}\varphi\bar\psi\psi
        +{\rm derivative\,\,coupling}\Big).             \label{1.6}
        \end{eqnarray}

This model is of a particular interest for a number of reasons. Firstly,
from the phenomenological viewpoint a strong nonminimal coupling allows
one to solve the problem of exceedingly small $\lambda$ (because here the
observable magnitude of CMBR anisotropy $\Delta T/T\sim 10^{-5}$ is
proportional to the ratio $\sqrt{\lambda}/|\xi|$ \cite{nonmin1,nonmin2}).
Secondly, this coupling is inevitable from the viewpoint of renormalization
theory. Also, among recent implications, it might be important in the theory
of an accelerating Universe \cite{BEPStar}. Finally, for a wide class of
GUT-type particle physics sectors this model generates a sharp probability
peak in $\rho_{\rm NB,T}(\varphi)$ at some $\varphi=\varphi_I $\cite{qsi,qcr}.
This peak belongs to the GUT energy scale -- the corresponding effective
Hubble constant, $H(\varphi_I)=\sqrt{\lambda/12|\xi|}\varphi_I$,
is proportional to $m_P\sqrt{\lambda}/|\xi|\sim 10^{-5}m_P$. This, in its
turn, justifies the use of GUT for matter part of the model, because
this scale is much below the supersymmetry and string theory scales.

The mechanism of this peak is based on a large
value of $|\xi|$ and the interaction (\ref{1.6}) which induces via the
Higgs effect large masses of all the particles directly coupled the inflaton.
Due to this effect the parameter $Z$ (dominated by terms quartic in particle
masses) is quadratic in $|\xi|$, $Z=6|\xi|^2\mbox{\boldmath$A$}/\lambda$,
with a universal combination of the coupling constants from (\ref{1.6})
        \begin{eqnarray}
        {\mbox{\boldmath $A$}} = \frac{1}{2\lambda}
        \Big(\sum_{\chi} \lambda_{\chi}^{2}
        + 16 \sum_{A} g_{A}^{4} - 16
        \sum_{\psi} f_{\psi}^{4}\Big).               \label{1.7}
        \end{eqnarray}

Therefore, the probability peak in this model reduces to the extremum of
the function
        \begin{eqnarray}
        \ln\rho_{NB,\,T}(\varphi)\simeq {\rm const}
        \pm\frac{24\pi(1+\delta)|\xi|}
        {\lambda}\frac{m_P^2}{\varphi^2}
        -3\frac{|\xi|^2}\lambda\mbox{\boldmath$A$}\,
        \ln\frac{\varphi^2}{\mu^2}
        +O\,\left(\frac{m_P^4}{\varphi^4}\right).       \label{1.8}
        \end{eqnarray}
Here the $\varphi$-dependent part of the classical instanton action is
taken in the lowest order of the slow roll smallness parameter,
$m_P^2/|\xi|\varphi^2\ll1$, renormalization ambiguous parameter $\mu$
enters only the overall normalization of $\rho_{\rm NB,\,T}(\varphi)$
and
        \begin{eqnarray}
        \delta\equiv
        -\frac{8\pi\,|\xi|\,m^2}{\lambda\,m_P^2}.    \label{delta}
        \end{eqnarray}

For the no-boundary and tunneling states the peak exists for positive
$\mbox{\boldmath$A$}$ and respectively negative and positive values of
$1+\delta$, $\pm(1+\delta)<0$.  The parameters of this peak -- mean values
of the inflaton and Hubble constants and the quantum width $\Delta$,
        \begin{eqnarray}
        &&\varphi_I= m_{P}\sqrt{\frac{8\pi|1+\delta|}{|\xi|
        {\mbox{\boldmath$A$}}}},\,\,\,\,\,
        H(\varphi_I)=
        m_{P}\frac{\sqrt{\lambda}}{|\xi|}
        \sqrt{\frac{2\pi|1+\delta|}
        {3{\mbox{\boldmath $A$}}}},             \label{1.9} \\
        &&\Delta=
        \frac 1{\sqrt{12{\mbox{\boldmath $A$}}}}
        \frac{\sqrt{\lambda}}{|\xi|}\varphi_I,    \label{1.10}
        \end{eqnarray}
are strongly suppresed by a small ratio $\sqrt{\lambda}/|\xi|$
known from the COBE normalization for $\Delta T/T\sim 10^{-5}$
\cite{COBE,Relict}. Because of small width the distribution function
can be approximated by the gaussian packet
        \begin{eqnarray}
        \rho_{NB,\,T}(\varphi)\simeq
        \frac1{\Delta\sqrt{2\pi}}
        \exp\left[-\frac{(\varphi-\varphi_I)^2}
        {2\Delta^2}\right].                            \label{1.11}
        \end{eqnarray}

The value of the parameter (\ref{delta}) is crucial for the inflationary
evolution from this gaussian peak. The classical equations of motion in the
slow roll approximation,
        \begin{eqnarray}
        &&\ddot{\varphi}+3H\dot{\varphi}-F(\varphi)=0,  \label{1.12}\\
        &&H(\varphi)\simeq\sqrt{\frac\lambda{12|\xi|}}\,\varphi,\,\,\,
        F(\varphi)\simeq-\frac{\lambda m_P^2
        (1+\delta)}{48\pi\xi^2}\,\varphi,          \label{1.13}
        \end{eqnarray}
show that the inflaton decreases from its initial value,
$\dot\varphi\simeq F/H<0$, only for $1+\delta>0$, that is only for the
{\em tunneling} state (minus sign in (\ref{1.8})). Only in this case
the duration of the inflationary epoch is finite with the e-folding number
(\ref{1.2}) \cite{efeq}
        \begin{eqnarray}
        N\simeq\frac{6\pi|\xi|\varphi_I^2}{m_P^2(1+\delta)}
        =\frac{48\pi^2}{\mbox{\boldmath $A$}}.   \label{1.14}
        \end{eqnarray}
Comparison with $N\geq 60$ immediately yields the
bound $\mbox{\boldmath $A$}\sim 5.5$ which can be regarded as a
selection criterion for particle physics models \cite{qsi}. This conclusion
remains qualitatively true when taking into account the contribution of
the inhomogeneous quantum modes to the radiation current of the
effective equations \cite{efeq}. This contribution and its dynamical effect
were obtained in \cite{efeq} by the method of the Euclidean effective action,
however, the quantum fluctuations of the inflaton field itself have not been
taken into account.

For the proponents of the no-boundary quantum states in a long debate on the
wavefunction discord \cite{VilVach,HawkTur2,VilLorentz,discorde} this
situation might seem unacceptable. According to this result the
no-boundary proposal does not generate realistic inflationary scenario,
while the tunneling state does not satisfy important aesthetic criterion
-- the universal formulation of the initial conditions and dynamical
aspects in one concept -- spacetime covariant path integral over
geometries\footnote{The Lorentzian path integral for the tunneling state
of \cite{VilLorentz} also requires, in this respect, extension beyond
minisuperspace level, development of the semiclassical expansion technique,
etc.}, not to say about intrinsic inconsistency mentioned in Introduction.
Thus, one of the motivations of considering the quantum mechanical
sector of the inflaton mode is the hope that it can handle this difficulty.
In view of the smallness of the quantum width (\ref{1.10}) the quantum
fluctuations $\Delta\varphi\sim\Delta$ are expected to be negligible, but
those of their quantum momenta $\Delta p_\varphi\sim 1/\Delta$ blow up
for small $\Delta$. Therefore, apriori, it is hard to predict the overall
magnitude of the quantum rolling force and its sign due to
$\Delta\varphi(t)$. In what follows we carefully consider this problem.

\section{Effective equations: setting the problem}
\hspace{\parindent}
Effective equations for expectation values of operators in the quantum
state $|{\mbox{\boldmath $\Psi$}}\big>$,
        \begin{eqnarray}
        &&g(x)=\big<{\mbox{\boldmath $\Psi$}}|
        \hat{g}(x)
        |{\mbox{\boldmath $\Psi$}}\big>,      \label{2.1}\\
        &&\hat{g}(x)=\hat{\varphi}(x),\hat{\chi}(x),
        \hat{\psi}(x),\hat{A}_\mu(x),\hat{g}_{\mu\nu}(x),..., \label{2.2}
        \end{eqnarray}
can be obtained by expanding the Heisenberg equations of motion,
$\delta S[\,\hat g\,]/\delta\hat g(x)=0$, for the full quantum field
$\hat{g}(x)=g(x)+\Delta\hat{g}(x)$ in powers of quantum disturbances
$\Delta\hat{g}(x)$
        \begin{eqnarray}
        &&\frac{\delta S[\,g\,]}{\delta g(x)}+\int dy\,
        \frac{\delta^2 S[\,g\,]}
        {\delta g(x)\,\delta g(y)}\,\Delta\hat g(y)  \nonumber\\
        &&\qquad\qquad+
        \frac12\int dy\,dz\,\frac{\delta^3 S[\,g\,]}
        {\delta g(x)\,\delta g(y)\,\delta g(z)}\,
        \Delta\hat g(y)\Delta\hat g(z)+...=0,       \label{2.2a}
        \end{eqnarray}
and averaging them with respect to $|{\mbox{\boldmath $\Psi$}}\big>$. The
linear in $\Delta\hat{g}(x)$ term identically drops out of this expansion,
because $\big<\Delta\hat{g}(x)\big>\equiv 0$, and the effective equations
acquire a generic form
        \begin{eqnarray}
        \frac{\delta S[\,g\,]}{\delta g(x)}
        +J(x)=0.                              \label{2.3}
        \end{eqnarray}
Here $S[\,g\,]$ is the classical action of the system, and the radiation
current $J(x)$ accumulates all quantum corrections which begin with the
one-loop contribution
        \begin{eqnarray}
        J(x)=\frac 12\int dy\,dz\,
        \frac{\delta^3 S[\,g\,]}
        {\delta g(x)\,\delta g(y)\,\delta g(z)}
        \,G(z,y)+...\,.                                 \label{2.4}
        \end{eqnarray}
The Wightman function of quantum disturbances $G(z,y)$ in a given quantum
state $|{\mbox{\boldmath $\Psi$}}\big>$
        \begin{eqnarray}
        &&G(z,y)=\big<{\mbox{\boldmath $\Psi$}}|\,\Delta\hat{g}(z)
        \,\Delta\hat{g}(y)\,
        |{\mbox{\boldmath $\Psi$}}\big>,          \label{2.5}\\
        &&\Delta\hat{g}(y)\equiv\hat{g}(y)-g(y),  \label{2.5a}
        \end{eqnarray}
can be found by iterations as a loop expansion in powers of $\hbar$. Because
semiclassically $\Delta\hat g=O(\hbar^{1/2})$ and $J(x)=O(\hbar)$, it follows
from the equations (\ref{2.2a}) and (\ref{2.3}) that the linear in
$\Delta\hat g$ term of (\ref{2.2a}) is at least linear in $\hbar$. Therefore,
in the one-loop approximation the quantum perturbation $\Delta\hat g(y)$
can be identified with the solution of the linearized classical equation on
the mean-field background
        \begin{eqnarray}
        \int dy\,\frac{\delta^2 S[\,g\,]}
        {\delta g(x)\,\delta g(y)}\,\Delta\hat g(y)=0.    \label{2.5b}
        \end{eqnarray}
Its solution can be parametrized by the operator-valued initial
conditions. Depending on the representation of the initial state
$|{\mbox{\boldmath $\Psi$}}\big>$, they can be either the
creation-annihilation operators, or operators of initial fields and their
conjugated momenta, so that quantum averaging in (\ref{2.5}) becomes
straightforward. Continuing this procedure by iterations one can obtain
the radiation current in any loop order as a complicated but, in
principle, calculable functional of the mean field.

Alternatively to (\ref{2.4})-(\ref{2.5}), the one-loop radiation current can
be written as
        \begin{eqnarray}
        J^{1-\rm loop}(x)=\frac 12\,
        \left<\,\left[\,\frac{\delta S[\hat g]}
        {\delta\hat g(x)}\,\right]_2 \right>,      \label{2.6}
        \end{eqnarray}
where $\big[...\big]_2$ denotes the quadratic part of the quantity expanded
in powers of disturbances $\Delta\hat g$ that solve linearized classical
equations, and $\big<...\big>$ implies the quantum averaging with respect to
$|{\mbox{\boldmath $\Psi$}}\big>$.

For the cosmological system this simple perturbative scheme should, however,
be amended by two important aspects. One of them reflects the local gauge
(general coordinate) invariance of the problem and the other deals with
disentangling the collective variables. The latter describes the most
important (minisuperspace) cosmological degrees of freedom having
non-trivial expectation values. In the next two sections we consider
these two aspects of the problem.

\subsection{Gauge properties of the radiation current}
\hspace{\parindent}
In view of local general coordinate and other gauge invariances, fields
and their perturbations contain purely gauge variables that should be
gauged away. Thus, the physical sector should be disentangled from
the full configuration space of the system and the physical state should be
prescribed on this physical sector. This is the general scheme of the
reduced phase space quantization \cite{BarvU,BKr,geom}. For describing
this scheme in application to perturbative radiation currents we simplify
the formalism by using
condensed notations for the full set of fields (\ref{2.1}), $g^a=g(x)$,
in which the condensed index $a$ includes both discrete spin labels
and spacetime coordinates $x$. Contraction of these indices implies
also the spacetime integration. In these notations, the invariance
of the action $S[g]$ under local gauge transformations,
$g^a\to g^a+R^a_\mu\,f^\mu$, with infinitesimal gauge parameters $f^\mu$
(the condensed index $\mu$ bearing together with discrete tensor labels
spacetime arguments of the local function $f^\mu=f(x)$) reads
        \begin{eqnarray}
        R^a_\mu\frac{\delta S}{\delta g^a}=0.          \label{g1}
        \end{eqnarray}
Here, $R^a_\mu$ is a generator of the gauge transformation -- the
quasilocal two-point kernel with respect to spacetime coordinates associated
with condensed labels $a$ and $\mu$.

A gauge breaking procedure -- a part of the physical reduction -- can be
enforced by adding to the classical action the gauge-breaking term and,
in the quantum domain -- for Heisenberg equations -- by including the
action of Faddeev-Popov ghosts. Then the derivation of effective equations
repeats the perturbative scheme of the above type with gauge-breaking and
ghost terms included into the full action. However, for the purpose of
physical reduction gauge conditions should be unitary, that is
transforming under gauge transformations locally in time (not involving
time derivatives of the gauge parameters $f^\mu$). In such a gauge the
ghosts are not propagating, and the ghost action can be omitted from the
total quantum system. As a result, the one-loop effective equations again
take the form (\ref{2.3}) with the same radiation current
        \begin{eqnarray}
        J_a^{\rm 1-loop}=\frac12 \frac{\delta^3 S}
        {\delta g^a\delta g^b\delta g^c}\,
        \big<\,\Delta\hat g^b\Delta\hat g^c\big>.        \label{g2}
        \end{eqnarray}
The only modification due to local invariance is that the linearized
equations of motion (\ref{2.5b}) for quantum perturbations
$\Delta\hat g^a$ are supplied with the linear gauge conditions
        \begin{eqnarray}
        &&\frac{\delta^2 S}{\delta g^a\delta g^b}\,\Delta\hat g^b=0,                \label{g3}\\
        &&\chi^\mu_a\,\Delta\hat g^a=0.            \label{g4}
        \end{eqnarray}
The functional matrix (two-point kernel) of the linear gauge condition
$\chi^\mu_a$ is assumed to form the Faddeev-Popov operator
        \begin{eqnarray}
        Q^\mu_\nu\equiv\chi^\mu_a\,R^a_\nu,         \label{g5}
        \end{eqnarray}
which is ultralocal in time, $Q^\mu_\nu\sim\delta(t_\mu-t_\nu)$ (the property
of the unitary gauge), and invertible. In view of this ultralocality the
inverse of $Q^\mu_\nu$, $Q^{-1\nu}_\mu$, does not require imposing any
boundary conditions in the past or future of the time variable.

The system of equations (\ref{g3})-(\ref{g4}) for quantum perturbations
$\Delta\hat g^b$ has a number of peculiarities. First, the linearized
gauge condition (\ref{g4}) guarantees that the gauge-breaking term
(usually quadratic in gauge conditions) does not contribute to the
radiation current (\ref{g2}). Second, it fixes uniquely the solution for
$\Delta\hat g^a$ under given initial conditions. In the absence of gauge
conditions, eq. (\ref{g3}) would have the ambiguity in the solution,
$\Delta\hat g^a\to\Delta\hat g^a+R^a_\mu\hat{f}^\mu$, with arbitrary
$\hat{f}^\mu$ because of a simple corollary of (\ref{g1})
        \begin{eqnarray}
        \frac{\delta^2 S}{\delta g^a\delta g^b}\,R^a_\mu=
        -\frac{\delta S}{\delta g^a}\,
        \frac{\delta R^a_\mu}{\delta g^b}.
        \end{eqnarray}
Here the right hand side vanishes on the classical solution,
$\delta S/\delta g^a=0$. So, the gauge generators $R^a_\mu$ are
zero vectors of the Hessian of the action on this background,
which implies the gauge invariance of the linearized
solution of the above type. However, the auxiliary condition
(\ref{g4}) gauges this invariance away and guarantees the uniqueness
of the solution for $\Delta\hat g^a$.

The parametrization of the general solution of eqs. (\ref{g3})-(\ref{g4}) in
terms of the symplectic (phase space) initial conditions is equivalent to
the Hamiltonian reduction of this linearized system to the physical sector.
This reduction should be done in the canonical formalism. The unitarity of
gauge conditions (\ref{g4}) guarantees that they can be rewritten in terms of the
phase space variables -- configuration coordinates and conjugated momenta --
contained in the set of $\Delta\hat g^a$ and $d\Delta\hat g^a/dt$ (the
rest of the variables in $\Delta\hat g^a$ are the Lagrange multipliers).
Solving these canonical
gauge conditions together with the first class constraints -- the
nondynamical subset of eq.(\ref{g3}) -- one finds all the perturbations
$\Delta\hat g^a$ as functions of the physical variables
$\Delta\hat g_{\rm phys}$ which in their turn become functions of initial
physical coordinates and momenta
$(\mbox{\boldmath$q$}_0,\mbox{\boldmath$p$}_0)$
        \begin{eqnarray}
        &&\Delta\hat g^a=
        \Delta\hat g^a(\Delta\hat g_{\rm phys}),          \label{g6}\\
        &&\Delta\hat g_{\rm phys}=
        \Delta\hat g_{\rm phys}
        (\mbox{\boldmath$q$}_0,\mbox{\boldmath$p$}_0).     \label{g7}
        \end{eqnarray}
If the initial quantum state is known on the physical sector in the
representation of these quantum variables,
$|\mbox{\boldmath$\Psi$}\big>=\mbox{\boldmath$\Psi$}(\mbox{\boldmath$q$}_0)$,
$\hat{\mbox{\boldmath$p$}}_0=\partial/i\partial\mbox{\boldmath$q$}_0$, then
the calculation of averages in (\ref{g2}) becomes straightforward.

In what follows we use this calculational strategy. The no-boundary and
tunneling wavefunctions as solutions of the Wheeler-DeWitt equations on
superspace will, first, be reduced to the physical sector. This quantum
reduction in the one-loop approximation will be performed by the technique of
\cite{BarvU,BKr,geom}\footnote{The fact that the cosmological states are
known as solutions of quantum Dirac constraints on superspace, and the
fact that their quantum reduction to the physical sector is readily
available from \cite{BarvU,BKr,geom}, explains why we work within unitary
gauge fixing procedure. Relativistic gauges with propagating
Faddeev-Popov ghosts would require a quantum state on extended Hilbert space
with indefinite metric, satisfying the zero BRST-charge equation (see
review of this problem in \cite{BarvU}). Lifting
the Dirac wavefunctions of the no-boundary and tunneling states to this
space, to the best of our knowledge, has not been done and goes beyond
the scope of this paper.}. Simultaneously,
the physical reduction will be performed for the cosmological
background and its perturbations, the both being parametrized in
terms of initial data encoded in the (reduced) physical wavefunction. This
makes further calculation of radiation currents straightforward.

We finish this section with gauge invariance of the radiation current --
the corollary of the Noether identity for the classical action (\ref{g1}).
The latter, after two consequitive functional differentiations, yields
another identity
        \begin{eqnarray}
        R^a_\mu\,\frac{\delta^3 S}
        {\delta g^a\delta g^b\delta g^c}
        =-\frac{\delta R^a_\mu}{\delta g^b}\,
        \frac{\delta^2 S}{\delta g^a\delta g^c}
        -\frac{\delta R^a_\mu}{\delta g^c}\,
        \frac{\delta^2 S}{\delta g^a\delta g^b}
        -\frac{\delta^2 R^a_\mu}{\delta g^b\delta g^c}\,
        \frac{\delta S}{\delta g^a}.                    \label{g8}
        \end{eqnarray}
Using it, one shows on account of the linearized equations
(\ref{g3}) that the radiation current satisfies the relation
        \begin{eqnarray}
        R^a_\mu J^{\rm 1-loop}_a=
        -\frac{\delta^2 R^a_\mu}{\delta g^b\delta g^c}\,
        \frac{\delta S}{\delta g^a}\,
        \big<\,\Delta\hat g^b\Delta\hat g^c\big>.          \label{g9}
        \end{eqnarray}
Here the right hand side vanishes on the classical background,
$\delta S/\delta g^a=0$, and, moreover, on an {\em arbitrary} mean field
background when the generator is linear in the field,
$\delta^2 R^a_\mu/\delta g^b\delta g^c=0$. But this is a well known property
of the generators of general coordinate transformations that
form the closed algebra of spacetime diffeomorphisms. Thus, the
one-loop radiation currents also satisfy the Noether identity
        \begin{eqnarray}
        R^a_\mu J^{\rm 1-loop}_a=0.                    \label{g10}
        \end{eqnarray}
This property will be very important in what follows. It implies that
the radiation currents are linearly dependent, which reduces the number of
effective equations to be solved in cosmological applications. Moreover, this identity
reflects the gauge invariance of effective equations themselves. In
particular, for purely gravitational system with $g^a(x)=g_{\mu\nu}(x)$, when
the radiation current coincides with the expectation value of the stress
tensor $J^{\mu\nu\, \rm 1-loop}(x)=\big<\hat{T}^{\mu\nu}(x)\big>$, this
property signifies the covariant conservation law,
$\nabla_\mu\big<\hat{T}^{\mu\nu}(x)\big>=0$.

\subsection{Two configuration space sectors of the model}
\hspace{\parindent}
In closed cosmological model, the total metric and inflaton scalar field
are usually decomposed into the spatially homogeneous
background and inhomogeneous perturbations
        \begin{eqnarray}
        &&ds^2=-N^2(t)dt^2+a^2(t)\gamma_{ij}dx^i dx^j
        +h_{\mu\nu}(x)dx^\mu dx^\nu,                   \label{2.7a}\\
        &&\varphi(x)=\varphi(t)+\delta\varphi(x), \,\,\,
        x=(t,\mbox{\bf x}),                            \label{2.7}
        \end{eqnarray}
where $a(t)$ is the scale factor, $N(t)$ is the lapse function and
$\gamma_{ij}$ is the spatial metric of the 3-sphere of unit radius.
Therefore, the full set of fields consists of the minisuperspace sector of
spatially homogeneous variables $Q(t)$ and inhomogeneous fields $f(x)$
essentially depending on spatial coordinates $x^i$={\bf x}
        \begin{eqnarray}
        &&g(x)=Q(t),\,f(t,\mbox{\bf x}),   \label{2.8a}\\
        &&Q(t)=a(t),\,\varphi(t),\,N(t),      \label{2.8b}\\
        &&f(x)=\delta\varphi(t,\mbox{\bf x}),
        \,h_{\mu\nu}(t,\mbox{\bf x}),
        \,\chi(t,\mbox{\bf x}),\,\psi(t,\mbox{\bf x}),\,
        A_\mu(t,\mbox{\bf x}),...                      \label{2.8}
        \end{eqnarray}
From the structure of the initial quantum state, that will be discussed later,
it follows that only minisuperspace variables have nonvanishing expectation
values
        \begin{eqnarray}
        \big<\,\hat{Q}(t)\,\big>\neq 0,\,\,\,\,
        \big<\,\hat{f}(x)\,\big>=0.                  \label{2.9}
        \end{eqnarray}
Therefore, the full set of effective equations reduces to the following three
equations in the minisuperspace sector
        \begin{eqnarray}
        &&\frac{\delta S\,[Q]}{\delta Q(t)}+J_Q(t)=0,   \nonumber\\
        &&J_Q=J_N,\,J_a,\,J_\varphi,                     \label{2.10}
        \end{eqnarray}
their quantum radiation currents $J_Q(t)$ containing the contribution of
quantum fluctuations of minisuperspace modes themselves and those of
spatially inhomogeneous fields.

The set of these equations is, however, redundant in view of the Noether
identities for both classical (\ref{g1}) and quantum (\ref{g10}) parts. In
the minisuperspace sector the general coordinate transformations reduce
to reparametrizations of time. Infinitesimal transformations of
minisuperspace variables $Q(t)=(N(t),\,a(t),\,\varphi(t))$, in the notations
of Sect. 3.1, read as: $f^\mu\equiv f(t),\,\nabla^a_\mu f^\mu\equiv
\nabla^Q f=(d(Nf)/dt,\,\dot a f,\,\dot\varphi f)$, and the identity
(\ref{g10}) takes the form
        \begin{eqnarray}
        \dot\varphi J_\varphi+\dot a J_a-N\dot J_N=0.   \label{2.10a}
        \end{eqnarray}
The currents $J_N$ and $J_\varphi$ have direct physical interpretation in
terms of the quantum energy density $\varepsilon$ and pressure $p$
        \begin{eqnarray}
        J_N=-a^3\sqrt{\gamma}\,\varepsilon,\,\,\,
        J_a=3a^2\sqrt{\gamma}\,p,            \label{2.11}
        \end{eqnarray}
so that eq.(\ref{2.10a}) in the cosmic time, $N=1$, takes the form
        \begin{eqnarray}
        \dot\varepsilon+3\,\frac{\dot a}a\,(\varepsilon+p)
        +\frac{J_\varphi}{\sqrt\gamma a^3}\,\dot\varphi=0.  \label{2.11a}
        \end{eqnarray}
As we shall see later, the third current $J_\varphi$ can be interpreted in
terms of the quantum rolling force driving the evolution of the inflaton
field. Therefore, this equation measures the balance of the conservation
for the quantum stress tensor vs the work of this force. In the slow roll
regime, $\dot\varphi\simeq 0$, it reduces to the conventional covariant
conservation law on the Robertson-Walker background.

The Hamiltonian reduction to the physical sector, discussed above,
        \begin{eqnarray}
        &&\Delta g(x)=(\Delta Q(t),\,f(x))\rightarrow
        \Delta g_{\rm phys}(x)=
        (\Delta Q_{\rm phys}(t),f_{\rm phys}(x)),  \\
        &&f_{\rm phys}(x)=
        (h^{\rm TT}(x), {\rm matter\,\,fields}),            \label{2.15}
        \end{eqnarray}
leaves us with the set of physical variables $\Delta g_{\rm phys}(x)$
which also splits into minisuperspace and field-theoretical subsets.
Here $\Delta Q_{\rm phys}(t)$ is a single field variable that originates
from the minisuperspace sector of metric and inflaton perturbations\footnote
{Counting the number of physical degrees of freedom is usual: in
3-dimensional configuration space of $(N,a,\varphi)$ subject to one
first class constraint the number of physical degrees of freedom equals
$3-2\times1=1$.}. The rest, $f_{\rm phys}(x)$, represent the transverse
traceless modes of the gravitational wave $h^{\rm TT}(x)$ and other
physical nongravitational degrees of freedom. The nature of
$\Delta Q_{\rm phys}(t)$ depends on the gauge used
for disentangling the physical sector. A particular gauge fixing procedure
widely used in the theory of cosmological perturbations \cite{GMTS,MukhAnd}
picks up a special gauge invariant variable
$\Delta Q_{\rm phys}(t)=\mbox{\boldmath$q$}(t)$ that will be discussed
in much detail in Sect.5. At the nonlinear level
(beyond perturbation theory on a classical background) another choice of
$Q_{\rm phys}(t)$ is possible by simply identifying it with the spatially
homogeneous inflaton field $\varphi(t)$, $Q_{\rm phys}(t)=\varphi(t)$. In
both cases, particular expressions for the original minisuperspace
variables $Q(t)$ and their perturbations $\Delta Q(t)$ in terms of
$Q_{\rm phys}(t)$ and $\Delta Q_{\rm phys}(t)$ depend on the choice
of gauge conditions. Two types of these gauge conditions will be considered
in Sects. 4 and 5.

Splitting the whole configuration space into minisuperspace and
inhomogeneous sectors (\ref{2.8a}) reflects the choice of the collective
variables. Moreover, it reflects distinctly different nature of
quantum states for the modes belonging to these two sectors. This
results in different calculational
strategies for the corresponding quantum averages. To see it, note that
on the space of physical variables $(\varphi,f_{\rm phys}(\mbox{\bf x}))$,
with $f_{\rm phys}(\mbox{\bf x})$ treated perturbatively, the initial
no-boundary and tunneling wavefunctions read (see Sect. 4)
        \begin{eqnarray}
        \mbox{\boldmath$\Psi$}^{\rm 1-loop}(\varphi,f_{\rm phys})=
        P(\varphi)\exp\left[\mp\frac12 I(\varphi)
        -\frac12\Omega(\varphi)f^2_{\rm phys}
        +O(f^3_{\rm phys})\right].          \label{2.16}
        \end{eqnarray}
Here $P(\varphi)$ is the loop prefactor and the tree-level exponential
contains the Euclidean action expanded up to a quadratic term in
inhomogeneous modes, $\Omega(\varphi)f^2_{\rm phys}$.

From (\ref{2.16}) it follows that the one-loop quantum correlators between
the minisuperspace modes and inhomogeneous fields vanish\footnote{Note that
due to the gaussian nature of the state $f\sim\hbar^{1/2}$, so that
the terms contributing to $\big<f\Delta\varphi\big>$-correlators,
$O(f^3)\sim\hbar^{3/2}$, go beyond the one-loop approximation.},
therefore they contribute additively to the total one-loop radiation current
        \begin{eqnarray}
        &&J=J^q+J^f,                          \label{2.12}\\
        &&J^q(t)=\frac 12\,
        \int dt'\,dt''\,
        \frac{\delta^3 S[\,Q\,]}
        {\delta Q(t)\,\delta Q(t')\,
        \delta Q(t'')}\,
        \big<\,\Delta\hat Q(t')
        \Delta\hat Q(t'')\,\big>,             \label{2.13}\\
        &&J^f(t)=\frac 12\,\int dx\,dy\,
        \left.\frac{\delta^3 S[\,Q+f\,]}
        {\delta Q(t)\,\delta f(x)\,
        \delta f(y)}
        \right|_{f=0}\!
        \big<\,\hat f(x)\hat f(y)\,\big>.     \label{2.14}
        \end{eqnarray}
Similarly to (\ref{2.6}) these contributions represent the quantum averages of
the quadratic terms of the expansion of $\delta S/\delta Q$ correspondingly
in $\Delta\hat Q(t)$ and $\hat f(x)$. However, the calculation methods for
$J^q$ and $J^f$ are very different, and the difference can be attributed to
qualitatively different quantum states of the modes $\varphi$ and
$f_{\rm phys}$. Let us begin with the radiation current $J^f$ which can
be obtained by the effective action method \cite{efeq}.

The matrix of quantum dispersions $\Omega(\varphi)$ in the gaussian state of
inhomogeneous modes is such that this state turns out to be the Euclidean
quasi-DeSitter invariant vacuum \cite{BAllen,Laf}
        \begin{eqnarray}
        |{\rm vac}\big>_{DS}=C(\varphi)
        \exp\left[-\frac12\Omega(\varphi)f^2_{\rm phys}\right].  \label{2.17}
        \end{eqnarray}
The corresponding quantum averages
        \begin{eqnarray}
        \vphantom{\big>}_{DS}\big<{\rm vac}|f_{\rm phys}(x)
        f_{\rm phys}(y)
        |{\rm vac}\big>_{DS}=G_{DS}(x,y)                   \label{2.18}
        \end{eqnarray}
are given by DeSitter invariant Green's functions which can be obtained by
analytic continuation from the unique Green's function on the Euclidean
section of the DeSitter spacetime -- the inverse of the Hessian of the
Euclidean action
        \begin{eqnarray}
        &&G_{DS}(x,y)=G_{E}(x_E,y_E)
        \Big|_{++++\,\rightarrow\,-+++},                 \label{2.19}\\
        &&\int dy_E\,\left.\frac{\delta^2\mbox{\boldmath$I$}[Q+f]}
        {\delta f(x_E)\,\delta f(y_E)}\right|_{f=0}
        G_{E}(y_E,z_E)=\delta(x_E,z_E),                   \label{2.20}\\
        &&\mbox{\boldmath$I$}[Q+f]=
        -iS[Q+f]\Big|_{++++\,\rightarrow\,-+++}.          \label{2.21}
        \end{eqnarray}
Here $x_E$ denotes the coordinates on the Euclidean DeSitter manifold
related by the analytic continuation to the Lorentzian
spacetime coordinates, $x^4_E=\pi/2H+ix^0,\,\mbox{\bf x}_E=\mbox{\bf x}$,
with $H$ -- the Hubble constant or the inverse radius of the Euclidean
4-sphere\footnote{The Lorentzian Green's function
$G_{DS}(x,y)=iG_{DS}^{(+)}(x,y)$ is the positive frequency Wightman function
-- the solution of the homogeneous linearized equation of motion.
On the contrary, $G_{E}(x_E,y_E)$, as an inverse of the Hessian, solves the
inhomogeneous equation. However, the Wightman function can be obtained from
$G_{E}(x_E,y_E)$ by taking the boundary value of its analytic continuation on
a proper shore of the cut in the complex plane of $[\sigma(x,y)]^{1/2}$ --
the geodetic distance between the points $x$ and $y$ \cite{BAllen}.} and
$x^0=t$ -- the cosmic time in the unperturbed DeSitter metric (\ref{2.7a})
corresponding to $N=1$, $a(t)=\cosh(Ht)/H$.

In view of this relation and in view of a similar analytic
continuation rule between the Lorentzian, $\delta^3 S/\delta Q\delta f^2$,
and Euclidean, $\delta^3\mbox{\boldmath$I$}/\delta Q\delta f^2$, 3-vertices,
one finds the {\em Euclidean effective action} algorithm for the radiation
current of inhomogeneous quantum modes
        \begin{eqnarray}
        &&J^f(x^0)=
        -\left.\frac{\delta\mbox{\boldmath$\SGamma$}_{\rm 1-loop}[\,Q\,]}
        {\delta Q(x^4)}
        \right|_{++++\,\rightarrow\,-+++},     \label{2.22}\\
        &&\mbox{\boldmath$\SGamma$}_{\rm1-loop}[\,Q\,]
        =\frac12\left.{\rm Tr\,ln}
        \frac{\delta^2\mbox{\boldmath$I$}[\,Q+f\,]}
        {\delta f(x_E)\,\delta f(y_E)}\right|_{f=0}.   \label{2.23}
        \end{eqnarray}
Note that this is exactly the functional that yields the one-loop
contribution to the distribution function (\ref{1.4}), when evaluated at
the classical solution for the minisuperspace background
$Q(x^4)=Q(x^4,\varphi)$ parametrically depending on $\varphi$,
        \begin{eqnarray}
        \mbox{\boldmath$\SGamma$}_{\rm1-loop}(\varphi)=
        \mbox{\boldmath$\SGamma$}_{\rm1-loop}
        [\,Q(x^4,\varphi)\,].  \label{2.24}
        \end{eqnarray}

This algorithm was used in the previous paper \cite{efeq} for the calculation
one-loop radiation currents of $f$-modes. In \cite{efeq}
$\mbox{\boldmath$\SGamma$}_{\rm1-loop}[\,Q\,]$ was obtained by the local
Schwinger-DeWitt expansion -- the expansion in spacetime derivatives of
the background fields, which in the
cosmological context corresponds to the slow-roll expansion\footnote{The
algorithm (\ref{2.22})-(\ref{2.23}) looks as a generalization of the
analytic continuation method for the effective equations in asymptotically
flat spacetime \cite{beyond}. Strictly speaking, this algorithm as derived
above holds only for {\em exact} DeSitter background, while the method of
\cite{beyond} was proven for arbitrary asymptotically flat backgrounds that
are perturbatively related to flat spacetime. For the deviations from the
DeSitter geometry (measured by the magnitude of the slow-roll smallness
parameter) the relations (\ref{2.22})-(\ref{2.23}) hold for local terms
of the effective action, and might not be true for an essentially nonlocal
part that cannot be expanded in powers of derivatives. But for the slow-roll
inflation regime, which we use throughout the paper, such an expansion --
the local Schwinger-DeWitt series -- is definitely applicable, which
justifies the effective action method.}. Within this expansion the one-loop
action is represented as a spacetime integral of the effective Lagrangian
expanded in powers of curvatures, matter field strengths and their
covariant derivatives. Therefore, the analytic continuation
rule (\ref{2.22}) is trivial -- the current $J^f(x^0)$ can be given by
the functional variation of the {\em local Lorentzian one-loop action}
        \begin{eqnarray}
        &&J^f(x^0)=
        \frac{\delta S_{\rm 1-loop}[\,Q\,]}
        {\delta Q(x^0)},                \label{2.24a}\\
        &&S_{\rm 1-loop}[\,Q(x^0)\,]=
        i\mbox{\boldmath$\SGamma$}_{\rm1-loop}[\,Q(x^4)\,]
        \Big|_{++++\,\rightarrow\,-+++}.
        \end{eqnarray}
Therefore, within the local expansion the $J^f$ part of the radiation current
can be absorbed in the functional variation of the total {\em Lorentzian}
effective action $S_{\rm eff}[\,Q\,]$, and the effective equations
acquire the final form
        \begin{eqnarray}
        &&\frac{\delta S_{\rm eff}[\,Q\,]}
        {\delta Q(x^0)}+J^q(x^0)=0,      \label{2.24b}\\
        &&S_{\rm eff}[\,Q\,]=
        S[\,Q\,]+S_{\rm 1-loop}[\,Q\,].    \label{2.24c}
        \end{eqnarray}
Here $S_{\rm eff}[\,Q\,]$ can be obtained from the classical action
$S[\,Q\,]$ by adding loop corrections to the classical coefficient
functions in the curvature and gradient expansion of the Lagrangian
\footnote{Note that the notion of
$S_{\rm eff}[\,Q\,]$, as a generator of equations for expectation
values, is legitimate only within the local derivative expansion. For
nonlocal contributions this action does not exist at all -- there is no
mean field functional that could yield by the variational procedure
effective equations for expectation values \cite{beyond}. The reason of this
is that for nonlocalities the analytic continuation from the Euclidean to
Lorentzian spacetime is not unique -- in addition, it requires setting
the retardation boundary conditions for nonlocal form factors
(see \cite{beyond} and cf. the previous footnote). These boundary conditions
prohibit the existence of the effective action for expectation values.}.
These corrections, in their turn, can be calculated in the Euclidean spacetime
by the local Schwinger-DeWitt technique \cite{BarvV}. For our model these
corrections has been obtained in \cite{efeq}. Thus, this essential
simplification in treating the $J^f$-part of the radiation current exists
due to the two important aspects of the problem -- DeSitter-invariant
vacuum of $f$-modes and the slow roll approximation. In contrast to this,
to the best of our knowledge, no simplification is available in the
calculation of the quantum mechanical part of the current $J^q$.

To begin with, the wavefunction of the quantum mechanical minisuperspace
mode $\varphi$ is not gaussian. Moreover, in the tree-level approximation the
graph of the probability distribution (\ref{1.3}) is very flat. It does not
have good probability peaks and is even unnormalizable. This means that
the tree-level quantum averages $\big<\Delta Q\Delta Q\big>^{\rm tree}$ are
badly defined. Beyond the tree-level approximation the situation can be
improved, because they should now be defined with the aid of the reduced
density matrix
        \begin{eqnarray}
        &&\big<\Delta Q(t)\Delta Q(t')\big>=
        {\rm tr}\left[\Delta\hat Q(t)\Delta\hat Q(t')\,
        \hat\rho\right],                          \label{2.25a}\\
        &&\hat\rho\equiv\rho(\varphi,\varphi')=\int df
        \mbox{\boldmath$\Psi$}(\varphi,f)
        \mbox{\boldmath$\Psi$}^*(\varphi',f),         \label{2.25}
        \end{eqnarray}
which originates from tracing the $f$-variables out and includes loop
corrections. As shown in \cite{norm,tunnel,BarvU,decoh}, the diagonal
element of this density matrix -- the distribution function of $\varphi$,
        \begin{eqnarray}
        \rho(\varphi)=\rho(\varphi,\varphi),         \label{2.26}
        \end{eqnarray}
is given in the approximation of a gaussian integral by the
effective action algorithm (\ref{1.4}). Effective action contributes the
factor that can generate a sharp probability peak (\ref{1.11}) with the
dispersion $\Delta$ defined by (\ref{1.10}). With this modification the
quantum correlators become well defined, being expressed in terms of
$\big<\Delta\varphi\Delta\varphi\big>\sim\Delta^2<\infty$. Certainly, this
improvement is achieved by exceeding the precision of the one-loop
approximation -- badly defined tree-level quantum correlators become
finite due to one-loop contribution (therefore, in their turn they
effectively contribute to the radiation currents two-loop quantities).
But overstepping the conventional rules of the loop expansion is
justified here because it reflects the underlying physics of the slow
roll dynamics.

Point is that the inflaton field in models satisfying the slow-roll
conditions effectively represents the massless scalar field -- its mass is
roughly proportional to the slow-roll smallness parameter \cite{qcr}. But
massless scalar fields do not have a well defined DeSitter invariant
vacuum \cite{BAllen}. This fact, in particular, manifests itself in the
unnormalizability of the tree-level wavefunction $\exp[\mp I(\varphi)/2]$,
absence of its local maxima, etc. As we see, loop effects render this
state a quasi-gaussian nature (\ref{1.11}) and thus justify the improved
semiclassical expansion. A major part of the paper in what follows deals with
the direct calculation of the quantum mechanical radiation current $J^q$
and its physical implications.

\section{Quantum Cauchy problem: tree level approximation}
\hspace{\parindent}
Loop expansion for effective equations is essentially perturbative. Therefore,
we solve them by iterations starting with the classical solution. Then, in
the one-loop approximation the radiation current can be calculated on the
classical background -- the lowest order approximation for the mean field.
Here we pose the initial conditions for this solution that follow from the
no-boundary and tunneling cosmological wavefunctions.

In this and the next section we work with the model of minimally
coupled inflaton field $\phi$ having a generic potential $V(\phi)$ (we
reserve the notation $\phi$ as opposed to the notation $\varphi$ for the
non-minimal inflaton). This general framework of the Cauchy problem
for the cosmological background and perturbations can be easily extended
to include the non-minimal model by reparametrizing the latter to the
Einstein frame \cite{open,renorm}, and this will be done in Sect. 6. Thus,
we begin with the action
        \begin{equation}
        S[g_{\mu\nu},\phi]=\int d^{4}x\, g^{1/2}
        \left(\frac{m_P^2}{16\pi}\,R(g_{\mu\nu})-
        \frac12(\nabla\phi)^2
        -V(\phi)\right).         \label{3.1}
        \end{equation}
Under the (unperturbed) ansatz for spatially homogeneous metric (\ref{2.7}),
it takes the minisuperspace form
        \begin{eqnarray}
        &&S[a,\phi,N]=\int dt\,Na^3\sqrt\gamma\left[\frac{3}{\kappa}
        \Big(\frac1{a^2}-\frac{\dot{a}^2}{N^2a^2}\Big)+
        \frac12\frac{\dot{\phi}^2}{N^2}-V(\phi)\right],     \label{3.3}\\
        &&\sqrt\gamma\equiv 2\pi^2,\,\,\,
        \kappa=\frac{8\pi}{m_P^2}.        \label{3.4}
        \end{eqnarray}
Classical equations for this action in the cosmic time gauge, $N=1$, read
        \begin{eqnarray}
        &&\frac1{a^3\sqrt\gamma}\,\frac{\delta S}
        {\delta N}\equiv\frac3\kappa\left(\frac1{a^2}
        +\frac{\dot a^2}{a^2}\right)
        -\frac{\dot\phi^2}2
        -V(\phi)=0,                                  \label{3.7a}\\
        &&\frac 1{Na^3\sqrt\gamma}\,\frac{\delta S}
        {\delta\phi}\equiv
        -\ddot\phi-3\,\frac{\dot a}a\,\dot\phi
        -V_\phi(\phi)=0,                             \label{3.7b}\\
        &&\frac 1{3Na^2\sqrt\gamma}\,\frac{\delta S}
        {\delta a}\equiv
        \frac1\kappa\left(\frac1{a^2}
        +2\,\frac{\ddot a}a
        +\frac{\dot a^2}{a^2}\right)
        +\frac{\dot\phi^2}2-V(\phi)=0,                \label{3.7c}
        \end{eqnarray}

The first of the eqs.(\ref{3.5}) represents the nondynamical Hamiltonian
constraint. In terms of the momenta conjugated to $a$ and $\phi$,
$\Pi_\phi=\sqrt\gamma a^3\dot\phi/N$, $\Pi_a=-6\sqrt\gamma a\dot a/\kappa N$,   \label{3.7}
this constraint has the following form
        \begin{eqnarray}
        \mbox{\boldmath$H$}(a,\phi,\Pi_a,\Pi_\phi)=
        -\frac\kappa{12a\sqrt\gamma}\Pi_a^2
        +\frac1{2a^3\sqrt\gamma}\Pi_\phi^2
        +a^3\sqrt\gamma\left[V(\phi)
        -\frac3{\kappa a^2}\right]=0,                \label{3.8}
        \end{eqnarray}
which at the quantum level in the coordinate representation of the quantum
minisuperspace, $\hat\Pi_a=\partial/i\partial a,\,\,\,
\hat\Pi_\phi=\partial/i\partial\phi$,
gives rise to the minisuperspace Wheeler-DeWitt equation on
$\mbox{\boldmath$\Psi$}(\phi,a)$
        \begin{eqnarray}
        \mbox{\boldmath$H$}(a,\phi,\partial/i\partial a,
        \partial/i\partial\phi)
        \mbox{\boldmath$\Psi$}(\phi,a)=0.       \label{3.10}
        \end{eqnarray}

There are two well known semiclassical solutions of this equation -- the
so-called no-boundary and tunneling wavefunctions. In the approximation of
the inflationary slow roll (when the derivatives with respect to $\phi$
are much smaller than the derivatives with respect to $a$) these two
solutions read \cite{VilVach}
        \begin{eqnarray}
        &&\mbox{\boldmath$\Psi$}_{NB}(\phi,a)=
        C_{NB}(a^2H^2(\phi)-1)^{-1/4}\exp\left[-\frac12 I(\phi)\right]
        \cos\left[S(a,\phi)+\frac\pi4\right],        \label{3.11}  \\
        &&\mbox{\boldmath$\Psi$}_{T}(\phi,a)=
        C_{T}(a^2H^2(\phi)-1)^{-1/4}\exp\left[+\frac12 I(\phi)
        +iS(a,\phi)+\frac{i\pi}4\right]                \label{3.12}
        \end{eqnarray}
They describe two types of the nucleation of the Lorentzian
quasi-DeSitter spacetime (described by the Hamilton-Jacobi function
$S(\phi,a)$) from the gravitational semi-instanton -- the Euclidean
signature hemisphere bearing the Euclidean gravitational action $I(\phi)/2$
        \begin{eqnarray}
        &&I(\phi)=-\frac{\pi m_P^2}{H^2(\phi)},    \nonumber \\
        &&S(\phi,a)=-\frac{\pi m_P^2}{2H^2(\phi)}
        (a^2H^2(\phi)-1)^{3/2}.                   \label{3.13}
        \end{eqnarray}
The size of this hemisphere -- its inverse radius -- as well as the
curvature of the quasi-DeSitter spacetime are determined by the effective
Hubble constant, $\dot{a}/a\simeq H(\phi)$, driving the inflationary dynamics
of the model
        \begin{eqnarray}
        H^2(\phi)=\frac{8\pi V(\phi)}{3m_P^2}=
        \frac{\kappa V(\phi)}3.                       \label{3.14}
        \end{eqnarray}

The nucleation of the Lorentzian spacetime from the Euclidean hemisphere
takes place at $a=1/H(\phi)$. This domain forms the one-dimensional curve
in the two-dimensional superspace. Its embedding equation can be written
in the form
        \begin{eqnarray}
        \chi(\phi,a)=a-\sqrt{\frac3{\kappa V(\phi)}}=0.   \label{3.15}
        \end{eqnarray}
The dimensionality of this subspace coincides with the number of
physical degrees of freedom in the minisuperspace sector of the model.
The intrinsic coordinate on this subspace becomes the physical
coordinate and the restriction of the Dirac wavefunction
$\mbox{\boldmath$\Psi$}(\phi,a)$ to this subspace
becomes the physical wavefunction, provided one takes care of a proper
relation between the quantum measures on the original superspace and the
physical subspace. For a generic constrained system, the details
of such a quantum reduction can be found in \cite{BarvU,BKr,geom} .
Here we just briefly repeat it for our model.

Let us identify $\chi(\phi,a)$ in (\ref{3.15}) with the gauge condition
fixing the time reparametrization invariance in the theory (\ref{3.3}) and
choose $\phi$ as the physical coordinate. Then,
according to the formalism of \cite{BKr}, the physical wavefunction
$\mbox{\boldmath$\Psi$}(\phi)$ in the one-loop (linear in
$\hbar$ approximation) can be obtained from the semiclassical Dirac
wavefunction $\mbox{\boldmath$\Psi$}(\phi,a)$ by the transformation
        \begin{eqnarray}
        \mbox{\boldmath$\Psi$}(\phi)=
        |\{\chi,\mbox{\boldmath$H$}\}|^{1/2}
        \mbox{\boldmath$\Psi$}(\phi,a)\Big|_{\chi(\phi,a)=0}. \label{3.16}
        \end{eqnarray}
Here we distinguish the original Dirac wavefunction in 2-dimensional
minisuperspace from the physical wavefunction by the number of their
arguments. The factor $|\{\chi,\mbox{\boldmath$H$}\}|$ -- the Poisson bracket
of the gauge condition with the first class Hamiltonian constraint --
is the Faddeev-Popov determinant
which should be calculated at the semiclassical values of momenta,
$\Pi_a=\partial_a S(\phi,a),\Pi_\phi=\partial_\phi S(\phi,a)$. In the slow
roll approximation, when the $\Pi_\phi$-momentum is negligible, this factor
equals $|\{\chi,\mbox{\boldmath$H$}\}|^{1/2}\sim(H^2a^2-1)^{1/4}$ and, thus,
cancels the preexponential factors in eqs.(\ref{3.11})-(\ref{3.12})
divergent at the nucleation surface (\ref{3.15}).

Thus, the physical wavefunction on the nucleation surface (\ref{3.15})
which should be regarded as {\em the Cauchy surface in minisuperspace}
reads as
        \begin{eqnarray}
        \mbox{\boldmath$\Psi$}_{NB,T}(\phi)=C_{NB,T}
        \exp\left[\mp\frac12 I(\phi)\right],          \label{3.17}
        \end{eqnarray}
minus and plus signs related respectively to the no-boundary and
tunneling states. It is well known that the graphs of these wavefunctions
are very flat for the situations when the slow roll approximation holds
(equivalent to small $\phi$-derivatives). Therefore, they
are generally not normalizable and do not have good probability peaks that
could be interpreted as a source of initial conditions for inflation.
The inclusion of loop terms via eq.(\ref{1.4}) might lead to the
normalizability of the wavefunction and, for the model of the nonminimally
coupled inflaton, even yield a sharp probability peak of the above type.
Then, the expectation value of the inflaton $\phi=\big<\hat\phi\big>$ becomes
finite. It is determined by the location of this peak and serves as the
initial condition for the classical extremal that will be used as the
background for the calculation of the one-loop radiation currents.

The second initial condition for this classical extremal
-- the time derivative of the inflaton -- arises from the expectation
value of the {\em physical} momentum conjugated to $\phi$,
$p_\phi=\big<{\hat p}_\phi\big>$. In view of reality of the initial
density matrix (\ref{2.25}) this expectation value is vanishing
        \begin{eqnarray}
        \big<\hat{p}_\phi\big>=\int d\phi\,
        \frac1{i}\frac\partial{\partial\phi}
        \rho(\phi,\phi')\Big|_{\phi'=\phi}=0.            \label{3.22}
        \end{eqnarray}
From the Hamiltonian reduction of the symplectic form in the gauge
$\chi(\phi,a)=0$ it follows that the physical momentum expresses in terms
of the original momenta
        \begin{eqnarray}
        &&\Pi_a da+\Pi_\phi d\phi
        =p_\phi d\phi,                               \nonumber\\
        &&p_\phi=\Pi_\phi-\Pi_a \chi_\phi/\chi_a,\,\,\,
        \chi_\phi\equiv\partial_\phi\chi,\,\,\,
        \chi_a\equiv\partial_a\chi.                  \label{3.23}
        \end{eqnarray}
Therefore, for $p_\phi=0$, $\Pi_\phi$ homogeneously expresses in terms of
$\Pi_a$ and, after plugging this relation into the Hamiltonian constraint
(\ref{3.8}), it implies that at the initial Cauchy surface $\Pi_\phi=0$
and $\Pi_a=0$. Thus, the full set of initial conditions for the classical
background reads
        \begin{eqnarray}
        \phi=\big<\hat\phi\big>,\,\,\,
        a=\frac1{H(\phi)},\,\,\,\dot\phi=\dot a=0.   \label{3.24}
        \end{eqnarray}

\section{Cauchy problem for cosmological perturbations}
\hspace{\parindent}
In this section we pose the Cauchy problem for quantum cosmological
perturbations propagating on the classical background of the previous
section. First, the set of perturbations is reduced by the technique
of \cite{GMTS} to the set of linearized invariants of spacetime
diffeomorphisms, and their quadratic action is constructed. The ghost nature
of their minisuperspace sector is revealed and the original perturbations
are built in terms of invariants in the Newton gauge. Then, quantum
initial conditions for perturbations are obtained with the aid
of the linearized version of the minisuperspace gauge introduced above.
Again, we consider the minimal model which will be later,
in Sect. 6, reparametrized to the non-minimal curvature coupling.

\subsection{Hamiltonian reduction to the physical sector}
\hspace{\parindent}
Here we start with the physical reduction for cosmological perturbations
on the classical background of Sect. 4. In the main, we follow the
notations of \cite{GMTS} where this reduction was presented in much detail.
In particular, we use the conformal time denoted by $\eta$ corresponding
to $N=a(\eta)$. In this gauge the classical equations of motion
(\ref{3.7a})-(\ref{3.7c}) have the form
        \begin{eqnarray}
        &&\frac3\kappa {\cal H}^2
        -\frac{\phi'^2}2+\frac3\kappa-a^2 V=0,    \nonumber\\
        &&{\cal H}^2+1-{\cal H}'=\frac\kappa2\phi'^2,    \nonumber \\
        &&\phi''+2{\cal H}\phi'+a^2 V_\phi=0,            \label{3.5}
        \end{eqnarray}
where primes denote the derivatives with respect to the conformal time,
subscript $\phi$ implies the partial derivative with respect to the inflaton,
$V_\phi\equiv \partial_\phi V(\phi)$, and ${\cal H}$ is the ``conformal''
Hubble constant
        \begin{eqnarray}
        {\cal H}\equiv\frac{a'}a,\,\,\,
        a'\equiv\frac{da}{d\eta},                     \label{3.6}
        \end{eqnarray}
related to the Hubble constant in cosmic time $H$ by the equation
${\cal H}=aH$.

The cosmological perturbations $(h_{ij},A,S_i,\delta\phi)$ of metric and
inflaton field are introduced according to the ansatz
        \begin{eqnarray}
        &&ds^2_{\rm total}=a^2(\eta)\left[-(1+2A)d\eta^2+2S_idx^id\eta
        +(\gamma_{ij}+h_{ij})dx^idx^j\right],             \label{4.1} \\
        &&\phi_{\rm total}=\phi+\delta\phi,        \label{4.2} \\
        &&h_{ij}=-2\psi \gamma_{ij}
        +2E_{|ij}+2F_{(i|j)}+t_{ij},               \label{4.3} \\
        &&S_i=\nabla_iB+V_i,\,\,\,\nabla_i F^i
        =\nabla_i V^i=t^i_i=\nabla^i t_{ij}=0.     \nonumber
        \end{eqnarray}
They consist of the scalar perturbations $(\psi,\delta\phi,E,A,B)$,
transverse vector perturbations $(F_i,V_i)$ and transverse-traceless tensor
ones $t_{ij}$. Here $\nabla_i $ denotes the spatial covariant derivative.

Spatially homogeneous modes from the minisuperspace sector, upon which
we focuse in this paper, belong to scalar perturbations. As discussed above,
the inhomogeneous modes which contribute to the $J^f$ radiation current can
be treated by the effective action method and do not require a manifest
physical reduction. Thus we consider only the scalar sector. After
constructing the quadratic part of the action in terms of scalar
perturbations one introduces the momenta conjugated to $(\psi,\delta\phi,E)$
        \begin{eqnarray}
        &&\Pi_\psi=\frac{2a^2\sqrt{\gamma}}\kappa\left[-3\Big(\psi'
        -\frac\kappa2\phi'\delta\phi
        +{\cal H}A\Big)-\Delta(B-E')\right],            \nonumber\\
        &&\Pi_{\delta\phi}=
        a^2\sqrt{\gamma}(\delta\phi'-\phi'A),            \nonumber\\
        &&\Pi_E=\frac{2a\sqrt{\gamma}}\kappa\Delta\left[\psi'
        -\frac\kappa2\phi'\delta\phi
        +{\cal H}A-(B-E')\right],                       \label{4.4}
        \end{eqnarray}
and finds out that $(A,B)$ play the role of Lagrange multipliers to the
linearized Hamiltonian and momentum constraints
        \begin{eqnarray}
        &&C_A=-{\cal H}\Pi_\psi+\phi'\Pi_{\delta\phi}
        + a^2\sqrt{\gamma}\left[-\frac2\kappa D\psi
        +({\cal H}\phi'-\phi'')\,\delta\phi\right],    \label{4.5}\\
        &&C_B=\Pi_E,                      \label{4.6}
        \end{eqnarray}
where $D$ is the following modified covariant Laplacian acting on a
closed 3-sphere with the metric $\gamma_{ij}$
        \begin{eqnarray}
        D=\Delta+3,\,\,\,
        \Delta=\gamma^{ij}\nabla_i\nabla_j.             \label{4.7}
        \end{eqnarray}

The constraints (\ref{4.5}) generate the diffeomorphisms in the scalar
perturbation sector with respect to the vector-field parameter
$\lambda^\mu=(\lambda^0,\nabla^i\lambda)$
        \begin{eqnarray}
        &&\delta_\lambda\psi=-{\cal H}\lambda^0,\,\,\,
        \delta_\lambda(\delta\phi)=\phi'\lambda^0,\,\,\,
        \delta_\lambda E=\lambda,                        \label{4.8}\\
        &&\delta_\lambda\Pi_\psi=
        \frac{2a^2\sqrt{\gamma}}\kappa D\lambda^0,\,\,\,
        \delta_\lambda\Pi_{\delta\phi}=
        a^2\sqrt{\gamma}(\phi''-{\cal H}\phi')\lambda^0,\,\,\,
        \delta_\lambda\Pi_E=0,                            \label{4.9}
        \end{eqnarray}
accompanied by the transformations of the Lagrange multipliers
        $\delta_\lambda A=(\lambda^0)'+{\cal H}\lambda^0$,
        $\delta_\lambda B=\lambda'-\lambda^0$.
There are two obvious invariants of the gauge canonical transformations
(\ref{4.8})-(\ref{4.9})
        \begin{eqnarray}
        &&\mbox{\boldmath$\Psi$}=\psi+
        \frac{\cal H}{\phi'}\delta\phi,               \label{4.11}\\
        &&\mbox{\boldmath$\Pi$}_\psi=\Pi_\psi-
        \frac{2a^2\sqrt{\gamma}}{\kappa\phi'}
        D\delta\phi.                                 \label{4.12}
        \end{eqnarray}
It turns out that after solving the constraints, $C_A=C_B=0$,
(\ref{4.5})-(\ref{4.6}), with respect to
$\Pi_{\delta\phi}$ and $\Pi_E$ and feeding the result into the canonical
action the latter entirely expresses in terms of these two invariants.
Moreover, they play the role of a single pair of canonically conjugated
variables in the physical sector \cite{GMTS}: on the constraint surface in
phase space the original symplectic form goes over into the physical one,
$\Pi_\psi\psi'+\Pi_{\delta\phi}\delta\phi'+\Pi_E E'=
\mbox{\boldmath$\Pi$}_\psi \mbox{\boldmath$\Psi$}'+(...)'$. The corresponding
canonical action quadratic in
$(\mbox{\boldmath$\Psi$},\mbox{\boldmath$\Pi$}_\psi)$ reads
        \begin{eqnarray}
        &&S[\mbox{\boldmath$\Psi$},\mbox{\boldmath$\Pi$}_\psi]
        \Big|_2=\int d\eta\,
        \left[\mbox{\boldmath$\Pi$}_\psi\mbox{\boldmath$\Psi$}'
        -\frac{2a^2\sqrt{\gamma}}{\kappa^2\phi'^2}
        \Big(D\mbox{\boldmath$\Psi$}
        +\frac{\kappa{\cal H}}{2 a^2 \sqrt{\gamma}}
        \mbox{\boldmath$\Pi$}_\psi\Big)^2\right.            \nonumber\\
        &&\qquad\qquad\qquad\quad\left.
        -\frac{a^2\sqrt\gamma}\kappa
        \mbox{\boldmath$\Psi$}D\mbox{\boldmath$\Psi$}
        +\frac{\kappa}{4a^2\sqrt\gamma}\mbox{\boldmath$\Pi$}_\psi
        \frac1D \mbox{\boldmath$\Pi$}_\psi\right]         \label{4.13}
        \end{eqnarray}
Somewhat simpler form this action acquires in terms of the new
variables $(\mbox{\boldmath$q$},\mbox{\boldmath$p$})$ related to
(\ref{4.11})-(\ref{4.12}) by the canonical transformation
        \begin{eqnarray}
        &&\mbox{\boldmath$q$}=\frac{2a}{\kappa\phi'}
        \mbox{\boldmath$\Psi$}
        +\frac{\cal H}{\phi'}\frac1{a\sqrt{\gamma}D}
        \mbox{\boldmath$\Pi$}_\psi,                      \nonumber\\
        &&\mbox{\boldmath$p$}=-\frac{\phi'a}{2\cal H}\sqrt{\gamma}
        D\mbox{\boldmath$\Psi$}
        +\frac{\kappa\phi'}{4a}
        \mbox{\boldmath$\Pi$}_\psi.                 \label{4.14}
        \end{eqnarray}
In terms of them the quadratic action in the {\em physical} sector of scalar
perturbations looks as
        \begin{eqnarray}
        &&S[\mbox{\boldmath$q$},\mbox{\boldmath$p$}]
        \Big|_2=\int d\eta\,
        \left[\mbox{\boldmath$p$}\mbox{\boldmath$q$}'+
        \mbox{\boldmath$p$}\mbox{\boldmath$q$}\left(\frac{\phi''}{\phi'}+
        \frac{\kappa\phi'^2}{4\cal H}\right)
        +\frac1{2\sqrt{\gamma}}\,\mbox{\boldmath$p$}\frac1{D}
        \mbox{\boldmath$p$}
        \right.                                      \nonumber\\
        &&\qquad\qquad\qquad\quad\left.
        +\frac{\kappa\phi'^2}{8{\cal H}^2}
        \left(-{\cal H}^2+1-\frac{\kappa\phi'^2}4\right)\,
        \sqrt{\gamma}\mbox{\boldmath$q$}D\mbox{\boldmath$q$}
        -\frac12\sqrt{\gamma}(D\mbox{\boldmath$q$})^2\right].  \label{4.15}
        \end{eqnarray}
With the extremal expression for the momentum
        \begin{eqnarray}
        \mbox{\boldmath$p$}=
        -\sqrt\gamma D\Big[\mbox{\boldmath$q$}'
        +(\phi''/\phi'+\kappa\phi'^2/4{\cal H})
        \mbox{\boldmath$q$}\Big]                       \label{4.15a}
        \end{eqnarray}
the Lagrangian form of this action is even shorter
        \begin{eqnarray}
        &&S[\mbox{\boldmath$q$}]\Big|_2
        =\frac12\int d\eta\,\sqrt{\gamma}(-D\mbox{\boldmath$q$})
        \left[-\frac{d^2}{d\eta^2}+\phi'\Big(\frac1{\phi'}\Big)''
        +\frac{\kappa\phi'^2}2+D\right]\mbox{\boldmath$q$}.   \label{4.16}
        \end{eqnarray}
The invariant field $\mbox{\boldmath$q$}$ here is well known from the theory
of cosmological perturbations \cite{GMTS}. It is actually given by the so
called Bardeen variable \cite{Bardeen,GMTS},
$\mbox{\boldmath$\Phi_H$}={\cal H}(B-E')-\psi,\,\,
\mbox{\boldmath$q$}=-2a\mbox{\boldmath$\Phi_H$}/\kappa\phi'$.

Note that the operator $D$ given by (\ref{4.7}) is negative definite except for
two modes: the zero mode corresponding to the Laplacian eigenvalue $\Delta=-3$
and the spatially homogeneous mode for $\Delta=0$, $D=+3$. In view of the
overall factor $-D$ the zero mode does not enter the action at all, while the
homogeneous mode enters (\ref{4.16}) with a wrong sign -- its kinetic term is
negative. Thus, this is a ghost variable signifying the classical instability
of the model. This instability at the linearized level is nothing but the
manifestation of the inflation which is a huge instability phenomenon
incorporating the runaway modes. In contrast with a conventional wisdom of the
S-matrix theory, this instability should not be regarded as an irrecoverable flaw
of the theory, because we know a nonlinear damping mechanism that provides
an exit from the inflation stage in case of the inflaton field rolling down
to smaller values of the potential. In particular, no special measures
like introducing the indefinite metric should be undertaken to eradicate
this phenomenon. Homogeneous fluctuations of the inflaton field do not have
a particle nature and one should not take care of guaranteeing the energy
positivity of their excitations. Therefore, this mode can and should be
quantized in the coordinate representation with positive metric in the
Hilbert space.

A single spatially homogeneous mode $\mbox{\boldmath$q$}(\eta)$ contained
in the full set of
        \begin{eqnarray}
        \mbox{\boldmath$q$}(x)=(\mbox{\boldmath$q$}(\eta),\,
        \mbox{\boldmath$q$}(\eta,\mbox{\bf x}))            \label{4.17}
        \end{eqnarray}
corresponds to the $D=+3$ eigenvalue of the operator (\ref{4.7}) in the action
(\ref{4.16}). It also satisfies all the above relations with a simple
ultralocal substitution $D=+3$ and is actually responsible for the
perturbations in the minisuperspace sector of the cosmological model. Indeed,
from the metric ansatz (\ref{4.1}), (\ref{4.3}) it follows that spatially
homogeneous variables $\psi(\eta)$, $\delta\phi(\eta)$ and $A(\eta)$ induced
by $\mbox{\boldmath$q$}(\eta)$ generate the variations of the scale factor,
inflaton field and lapse function
        \begin{eqnarray}
        \delta a=-a\psi+O(\psi^2),\,\,\,\delta\phi,\,\,\,
        \delta N=aA+O(A^2).                                \label{4.18}
        \end{eqnarray}

Actual expression for $\psi(\eta)$, $\delta\phi(\eta)$ and $A(\eta)$ in terms
of $(\mbox{\boldmath$q$}(\eta),\mbox{\boldmath$p$}(\eta))$ depend on the
particular gauge chosen for minisuperspace variables. In what follows we will
need two types of such gauges. One will be used for gauge fixing the
dynamical evolution of perturbations as a function of dynamically evolving
invariant variable $\mbox{\boldmath$q$}(\eta)$. Another gauge serves as a part
of the quantum Cauchy problem -- as shown in the previous section, it
facilitates the quantum reduction to the physical sector and relates
the wavefunction to the initial conditions for both the classical
background and the {\em homogeneous} perturbation
variable $\mbox{\boldmath$q$}(\eta)$. The first gauge may coincide with the
second one. However, its use is strongly biased by practical necessities of
the theory of cosmological perturbations \cite{BFM} and, therefore, is
usually chosen to be the Newton gauge which is essentially different from the
minisuperspace gauge of Sect. 3. Thus we consider these two gauges
separately.

\subsection{Newton gauge}
\hspace{\parindent}
Newton gauge is widely used in the theory of cosmological perturbations
\cite{BFM} to express them in terms of the Bardeen invariant
$\mbox{\boldmath$q$}$. The Newton gauge for spatially inhomogeneous modes
reads
        \begin{eqnarray}
        B=0,\,\,\,E=0.                         \label{4.19}
        \end{eqnarray}
From the equations for momenta (\ref{4.4}) and the momentum constraint $C_B=0$
this implies that
        \begin{eqnarray}
        \Pi_\psi=-\frac{6 a^2\sqrt{\gamma}}\kappa\Big(\psi'
        -\frac\kappa2\phi'\delta\phi+{\cal H}A\Big)=0.     \label{4.20}
        \end{eqnarray}
In the spatially homogeneous sector of the theory, where the contribution of
$B$ and $E$ is missing (they enter only differentiated with respect to spatial
coordinates) the latter equation should be regarded as the definition of the
Newton gauge. This gauge involves only the phase space coordinate -- the
momentum $\Pi_\psi$ -- and, therefore, it is unitary and can be identified
with the gauge (\ref{g4}).

Canonical equations of motion for
$(\mbox{\boldmath$\Psi$},\mbox{\boldmath$\Pi$}_\psi)$, which follow from the
action (\ref{4.13}), in this gauge have a simple corollary
$(a\psi)'/a-\kappa\phi'\delta\phi/2=0$. When compared with (\ref{4.4}),
$\Pi_\psi=0$, this
corollary yields the main relation in the Newton gauge
        \begin{eqnarray}
        A=\psi.                             \label{4.21}
        \end{eqnarray}
Then one easily expresses all the perturbations in terms of the physical
phase space variables $(\mbox{\boldmath$q$},\mbox{\boldmath$p$})$. On
substituting the Lagrangian value of the momentum (\ref{4.15a}) these
expressions finally simplify to
        \begin{eqnarray}
        &&\psi=\frac{\kappa\phi'}{2a}\mbox{\boldmath$q$}, \\
        &&\delta\phi=
        \frac{(\phi'\mbox{\boldmath$q$})'}{a\phi'}.       \label{4.22}
        \end{eqnarray}
Eqs. (\ref{4.21})-(\ref{4.22}) form the needed set of relations (\ref{g6})
of the physical reduction for minisuperspace perturbations.

\subsection{Minisuperspace gauge}
\hspace{\parindent}
The minisuperspace gauge $ \chi(a,\phi)=0$ of Sect. 3 was used for the
physical reduction of the minisuperspace wavefunction and for establishing
the tree-level initial conditions -- for the classical background. Let us now
use it in order to find the initial conditions for
$(\mbox{\boldmath$q$},\mbox{\boldmath$p$})$. The linearized minisuperspace
gauge condition (the gauge (\ref{g4}) in condensed notations of Sect. 3.1)
gives the perturbation of $a$ in terms of $\delta\phi$. Taking into
account the relations (\ref{4.18}), expressing $\delta a$ in terms of the
perturbation $\psi$, we get
        \begin{eqnarray}
        \psi=\frac{\chi_\phi}{a\chi_a}\delta\phi.         \label{4.23}
        \end{eqnarray}
The corresponding reduction of the symplectic form gives the expression
for the physical momentum $p_{\delta\phi}$ conjugated to $\delta\phi$
        \begin{eqnarray}
        \Pi_\psi\psi'+\Pi_{\delta\phi}\delta\phi'
        =p_{\delta\phi}\delta\phi'+...,\,\,\,\,
        p_{\delta\phi}=\Pi_{\delta\phi}
        +\Pi_\psi\frac{\chi_\phi}{a\chi_a}.        \label{4.24}
        \end{eqnarray}
Then, by solving the linearized constraint $C_A=0$ one easily finds
$\Pi_\psi$ and $\Pi_{\delta\phi}$ as functions of
$(\delta\phi,p_{\delta\phi})$ and, via the formalism above, proceeds to the
final transformation relating $(\mbox{\boldmath$q$},\mbox{\boldmath$p$})$
to $(\delta\phi,p_{\delta\phi})$
        \begin{eqnarray}
        &&\mbox{\boldmath$q$}(\eta)=\frac1{M}
        \left[\Big({\cal H}-\frac{\phi''}{\phi'}\Big)
        \frac{a^2\chi_a{\cal H}}3
        +\frac{2\chi^2_\phi}{\kappa\chi_a}\right]\delta\phi(\eta)
        +\frac{\chi_a{\cal H}}{3M\sqrt\gamma}
        p_{\delta\phi}(\eta),                           \label{4.25}\\
        &&\mbox{\boldmath$p$}(\eta)=
        \sqrt\gamma\left\{\frac{\kappa\phi'^2}{4M}
        \left[\Big({\cal H}-\frac{\phi''}{\phi'}\Big)
        \frac{a^2\chi_a{\cal H}}3
        +\frac{2\chi^2_\phi}{\kappa\chi_a}\right]
        -\frac M{\chi_a}\right\}
        \frac 3{\cal H}\delta\phi(\eta)
        +\frac{\chi_a}M\frac{\kappa\phi'^2}4
        p_{\delta\phi}(\eta),                          \label{4.26}\\
        &&M\equiv\phi'\chi_\phi
        +{\cal H}a\chi_a.                            \label{4.27}
        \end{eqnarray}
Note that this relation is written down in the homogeneous sector which is
emphasized by the time arguments of the phase space variables. In the
right hand side of these equations the spatial homogeneity manifests itself
in the particular value of the operator $D$, $D=3$. One can easily check
that this transformation is canonical,
$\{\mbox{\boldmath$q$},\mbox{\boldmath$p$}\}=\{\delta\phi,p_{\delta\phi}\}=1$,
and invertible. Inverting it, one can find all the minisuperspace
perturbations as functions of $\mbox{\boldmath$q$}$ and
$\mbox{\boldmath$q$}'$, similarly to the relations (\ref{4.21})-(\ref{4.22})
in the Newton gauge. However, the goal of working in the gauge (\ref{3.15})
is somewhat different. We shall need the relations (\ref{4.25})-(\ref{4.27})
in order to express initial conditions
for $(\mbox{\boldmath$q$}(\eta),\mbox{\boldmath$p$}(\eta))$ in terms of
initial conditions for $(\delta\phi,p_{\delta\phi})$. The latter in their
turn follow from the cosmological wavefunction in the physical sector.

We begin by noting that at the initial moment of time the following relations
hold
        \begin{eqnarray}
        &&{\cal H}=\eta+O(\eta^2),\,\,\,
        \phi'=-\frac{3V_\phi}{\kappa V}\eta+O(\eta^2), \,\,\,
        \left.\frac{\cal H}{\phi'}\right|_0=
        \left.\frac{\cal H'}{\phi''}\right|_0=
        -\frac{\kappa V}{3V_\phi},             \nonumber\\
        &&M=\eta\Big(\frac3{\kappa V}\Big)^{1/2}
        \left(1-\frac32\frac{V_\phi^2}{\kappa V^2}\right)+O(\eta^2),\,\,\,
        \eta\rightarrow 0.         \nonumber
        \end{eqnarray}
Using these relations in eqs(\ref{4.25})-(\ref{4.26}) we obtain the asymptotic
behavior of the invariant variables
$(\mbox{\boldmath$q$},\mbox{\boldmath$p$})$ for $\eta\to 0$ in terms of
the physical variables of the minisuperspace gauge fixing
$(\delta\phi(\eta),p_{\delta\phi}(\eta))$
        \begin{eqnarray}
        &&\left(\frac{\kappa V}3\right)^{1/2}
        \mbox{\boldmath$q$}\simeq-\frac1{3\eta}
        \delta\phi+\frac{\kappa V}9
        \frac1{1-9\epsilon^2/16}
        \frac{p_{\delta\phi}}{\sqrt\gamma},       \label{4.28} \\
        &&\left(\frac3{\kappa V}\right)^{1/2}
        \mbox{\boldmath$p$}
        \simeq-\frac9{\kappa V}(1-9\epsilon^2/32)
        \sqrt\gamma\delta\phi,\,\,\,\,\,
        \eta\rightarrow 0.                       \label{4.29}
        \end{eqnarray}
Just to emphasize the role of the slow roll expansion we retained here the
corrections proportional to the smallness parameter
        \begin{eqnarray}
        \epsilon^2\equiv\frac83
        \frac{V_\phi^2}{\kappa V^2}\ll 1.   \label{4.30}
        \end{eqnarray}
In what follows we shall systematically discard such corrections retaining only
the leading order of the slow roll expansion.

Important peculiarity of the behavior (\ref{4.28})-(\ref{4.29}) is its
singularity at $\eta\to 0$. This singularity is, however, an artifact of the
definition of the invariant variables (\ref{4.14}) nonanalytic at
$\phi'\to 0$, rather than the manifestation of some physical inconsistencies.
To see it, one can decompose the general classical solution for
$\mbox{\boldmath$q$}(\eta)$ in the sum of two linearly independent solutions
of the equation of motion for the action (\ref{4.16})
        \begin{eqnarray}
        &&\mbox{\boldmath$q$}(\eta)
        =c_+\mbox{\boldmath$q$}_+(\eta)+
        c_-\mbox{\boldmath$q$}_-(\eta),    \label{4.31a}\\
        &&\left(-\frac{d^2}{d\eta^2}
        +\phi'\Big(\frac1{\phi'}\Big)''
        +\frac{\kappa\phi'^2}2+3\right)
        \mbox{\boldmath$q$}_\pm(\eta)=0.        \label{4.31}
        \end{eqnarray}
Because of $\phi'(1/\phi')''\sim 2/\eta^2,\,\,\eta\to 0$, the initial
moment $\eta=0$ is a singular point of this differential
equation, at which one of the two solutions, $\mbox{\boldmath$q$}_-(\eta)$,
diverges as $1/\phi'$. One can make a singular rescaling,
        \begin{eqnarray}
        \mbox{\boldmath$q$}\equiv
        \frac{\mbox{\boldmath$Q$}}{\phi'},           \label{4.32}
        \end{eqnarray}
to a new variable $\mbox{\boldmath$Q$}(\eta)$ which is finite at this point.
It satisfies the equation
        \begin{eqnarray}
        \left(-\frac{d^2}{d\eta^2}+2\frac{\phi''}{\phi'}
        \frac d{d\eta}
        +\frac{\kappa\phi'^2}2+3\right)
        \mbox{\boldmath$Q$}(\eta)=0,                 \label{4.33}
        \end{eqnarray}
and has as two solutions the following regular functions
        \begin{eqnarray}
        &&\mbox{\boldmath$Q$}_+(\eta)=
        \eta^3\left(1+O(\eta^2)\right),               \label{4.34} \\
        &&\mbox{\boldmath$Q$}_-(\eta)=
        1-3\eta^2/2+O(\eta^3).                         \label{4.35}
        \end{eqnarray}

Substituting the decomposition (\ref{4.31a}), with $\mbox{\boldmath$q$}_\pm$
related to $\mbox{\boldmath$Q$}_\pm$ by (\ref{4.32}), to the left hand sides
of (\ref{4.28})-(\ref{4.29}) one obtains the system of equations for
$c_{\pm}$ with singular coefficients. This system, however, has a regular
solution in terms of the initial conditions for physical variables
$(\delta\phi(0),p_{\delta\phi}(0))$
        \begin{eqnarray}
        &&c_+=-\frac13\left(\frac{\kappa V}3\right)^{1/2}
        \frac{V_\phi}{\kappa V}\,
        \frac{p_{\delta\phi}(0)}{\sqrt\gamma},       \label{4.36}\\
        &&c_-=\left(\frac{3}{\kappa V}\right)^{1/2}
        \frac{V_\phi}{\kappa V}
        \,\delta\phi(0).                           \label{4.37}
        \end{eqnarray}
This basic relation will be used throughout the rest of the paper to
express the Heisenberg operators of quantum perturbations
$\Delta\hat Q_{\rm phys}(\eta)$ and $\Delta\hat Q(\eta)$ in terms of
the Schrodinger operators,
$\delta\hat\phi(0)=\delta\phi,\,
\hat p_{\delta\phi}(0)=\partial/i\partial(\delta\phi)$, and then find the
quantum averages of their bilinear combinations in the inflaton
representation of the initial density matrix (\ref{2.25}).

\section{Non-minimal model}
\hspace{\parindent}
In what follows we go over to the model (\ref{1.5}) that has a
good peak-like behavior of the initial distribution function of the inflaton
\cite{qsi,qcr,tvsnb}. The inflaton-graviton sector of the action in this
model can be rewritten in the form
        \begin{eqnarray}
        S[g_{\mu\nu},\varphi]=\int d^4x\, g^{1/2}\left\{-V(\varphi)
        +U(\varphi)R
        -\frac12 G(\varphi)(\nabla\varphi)^2\right\}.       \label{5.1}
        \end{eqnarray}
In fact, the curvature (and derivative) expansion of any low-energy effective
graviton-scalar action can be truncated to this form with some coefficient
functions of the zeroth and first order in the curvature -- the scalar field
potential $V(\varphi)$, the effective $\varphi$-dependent Planck
``mass'' $16\pi U(\varphi)$ and the coefficient of the inflaton kinetic
term $G(\varphi)$. In the classical model (\ref{1.5}) these functions have a
particular form
        \begin{eqnarray}
        &&U(\varphi)=\frac{m_P^2}{16\pi}
        +\frac12 |\xi|\,\varphi^2,                    \label{5.6}\\
        &&V(\varphi)=\frac{m^2\varphi^2}2
        +\frac{\lambda\varphi^4}4,                    \label{5.6a}\\
        &&G(\varphi)=1.                               \label{5.6b}
        \end{eqnarray}

It is well known that the action (\ref{5.1}) can be transformed to
the Einstein frame by a special conformal transformation and
reparametrization of the inflaton field
$(g_{\mu\nu},\varphi)\rightarrow(\bar g_{\mu\nu},\bar\varphi)$,
        \begin{eqnarray}
        &&S[g_{\mu\nu},\varphi]=
        \bar S[\bar g_{\mu\nu},\bar\varphi],       \nonumber\\
        &&\bar S[\bar g_{\mu\nu},\bar\varphi]=
        \int d^4x\,\bar g^{1/2}\left\{-\bar V(\bar\varphi)
        +\frac{m_P^2}{16\pi}R(\bar g_{\mu\nu})
        -\frac12(\bar\nabla\bar\varphi)^2\right\}.   \label{5.4}
        \end{eqnarray}
In what follows, we shall denote the fields and other objects in the
Einstein frame of the non-minimal model by bars and identify them with
those of the minimal model considered in Sects. 3 - 5. In this way we
reduce all the calculations, Cauchy data setting, gauge fixing, reduction
to the physical sector, etc. to the algorithms derived above for the case
of the minimal model.

\subsection{Reparametrization to the minimal frame}
\hspace{\parindent}
The transformations relating the actions $S[g_{\mu\nu},\varphi]$ and
$\bar S[\bar g_{\mu\nu},\bar\varphi]$ are implicitly given by equations
\cite{renorm,open}
        \begin{eqnarray}
        &&\bar g_{\mu\nu}=\frac{16\pi U(\varphi)}
        {m_P^2}g_{\mu\nu},                            \label{5.2}\\
        &&\left(\frac{d\bar\varphi}{d\varphi}\right)^2
        =\frac{m_P^2}{16\pi}\frac{GU+3U_\varphi^2}{U^2},      \label{5.3}
        \end{eqnarray}
where, similarly to previous sections, $\varphi$-subscripts denote the
derivatives of the coefficient functions with respect to the inflaton,
$V_\varphi\equiv dV/d\varphi,\,V_{\varphi\varphi}\equiv d^2V/d\varphi^2$,
etc. The action in terms of new fields (\ref{5.4}) has a minimal coupling
and the new inflaton potential
        \begin{eqnarray}
        \bar{V}(\bar\varphi)=\left.\left(\frac{m_P^2}{16\pi}\right)^2
        \frac{V(\varphi)}{U^2(\varphi)}
        \,\right|_{\varphi=\varphi(\bar\varphi)}.            \label{5.5}
        \end{eqnarray}
For the coefficient functions (\ref{5.6})-(\ref{5.6a}) the explicit
reparametrization between the frames can be found for large value of
the nonminimal coupling constant $|\xi|\gg 1$ and small value of the
parameter $m_P^2/|\xi|\varphi^2\ll 1$ \cite{open}
        \begin{eqnarray}
        &&\varphi(\bar\varphi)\simeq\frac{m_P}{|\xi|^{1/2}}
        \exp\left[\sqrt{4\pi/3}
        \Big(1+\frac1{6\,|\xi|}\Big)^{-1/2}
        \frac{\bar\varphi}{m_P}\right],                 \label{5.7}\\
        &&\bar{V}(\bar\varphi)=\frac{\lambda m_P^4}{256\pi^2|\xi|^2}
        \,\left[\,1-\frac{1+\delta}{4\pi}\frac{m_P^2}
        {|\xi|\,\varphi^2}+...\right]
        _{\,\varphi=\varphi(\bar\varphi)},                  \label{5.8}
        \end{eqnarray}
where we have retained only the first order term in $m_P^2/|\xi|\,\varphi^2$.
In view of (\ref{5.7}), for large $\bar\varphi$ this potential exponentially
approaches a constant and satisfies the slow roll approximation with the
expansion parameter \cite{open}
        \begin{eqnarray}
        \epsilon=\frac{m_P}{\sqrt{3\pi}}
        \frac{\bar V_{\bar\varphi}(\bar\varphi)}
        {\bar V(\bar\varphi)}
        \simeq\frac{1+\delta}{3\pi}
        \left(1+\frac1{6\,|\xi|}\right)^{-1/2}
        \frac{m_P^2}{|\xi|\,\varphi^2}\ll 1,          \label{5.9}
        \end{eqnarray}
which justifies the smallness of the parameter $m_P^2/|\xi|\varphi^2$ chosen
above.

Let us now consider the minimal model of Sects. 3 - 5 as an Einstein frame
of the non-minimal model and label all the objects of the minimal model --
the metric, inflaton field, scale factor, conformal time, cosmological
perturbations and the minisuperspace gauge fixing conditions -- by bars
        \begin{eqnarray}
        &&\bar{g}_{\mu\nu},\,\,\phi=\bar\varphi,\,\,
        \bar a,\,\,\bar\eta,                               \nonumber\\
        &&\bar\psi,\,\, \delta\phi=\delta\bar\varphi,\,\,
        \bar A,\,\,\bar\Pi_\psi,\,\,\bar\Pi_{\delta\phi},\,\,
        \bar\chi(\bar a,\bar\varphi),
        \end{eqnarray}
as opposed to the objects in the original -- non-minimal -- frame:
$g_{\mu\nu},\varphi,a,\eta,\psi,\delta\varphi,A,\chi(a,\varphi)$.
Comparing the metrics in these frames, perturbed by the cosmological
disturbances from the scalar sector,
        \begin{eqnarray}
        &&ds^2=a^2\left[-(1+2A)d\eta^2
        +(1-2\psi)\gamma_{ij}dx^idx^j\right],        \nonumber\\
        &&d\bar s^2=\bar a^2\left[-(1+2\bar A)d\bar\eta^2
        +(1-2\bar\psi)\gamma_{ij}dx^idx^j\right],     \label{5.10}\\
        &&d\bar s^2=\frac{16\pi U}{m_P^2} ds^2,
        \end{eqnarray}
one finds the relations between these two sets of variables
        \begin{eqnarray}
        &&\bar a=\sqrt{\frac{16\pi
        U(\varphi)}{m_P^2}}a\simeq
        \sqrt{\frac{8\pi|\xi|\varphi^2}{m_P^2}}a,\,\,\,
        \bar\eta=\eta,                              \label{5.14}\\
        &&\delta\phi\equiv\delta\bar\varphi
        =\sqrt{\frac{m_P^2}{16\pi}\frac{U+3U'^2}{U^2}}\simeq
        \sqrt{\frac3{4\pi}}\frac{m_P}
        \varphi \delta\varphi,                        \label{5.11} \\
        &&\bar\psi=\psi-\frac{U_\varphi}{2U}
        \delta\varphi
        \simeq\psi-\frac{\delta\varphi}\varphi,      \label{5.12}\\
        &&\bar A=A+\frac{U_\varphi}{2U}\delta\varphi
        \simeq A+\frac{\delta\varphi}\varphi,         \label{5.13}
        \end{eqnarray}
where the last three relations hold in the linear order of perturbation theory
in cosmological disturbances. The canonical momenta $\Pi_{\delta\varphi}$
and $\Pi_\psi$ obviously transform by the rule contragradient to
(\ref{5.11}) and (\ref{5.12}).

We also need the frame transformation between the physical sectors
defined in the minisuperspace gauge of Sect. 5.3. The gauge condition
itself transforms as a scalar -- only in this case it represents one
and the same Cauchy surface, written in two different coordinate
systems on minisuperspace
        \begin{eqnarray}
        \bar\chi(\bar a,\bar\varphi)=\chi(a,\varphi).    \label{5.19}
        \end{eqnarray}
As regards the reparametrization beteween these coordinate systems,
(\ref{5.7}) and (\ref{5.14}), it has a general form
        \begin{eqnarray}
        \bar\varphi=\bar\varphi(\varphi),\,\,\,
        \bar a=\bar a(\varphi,a),         \label{5.16}
        \end{eqnarray}
mixing the inflaton and the scale factor only in the transformation of $a$.
Therefore, the linearized perturbations of $a$ and $\varphi$ and their momenta
in both frames are related by a triangular transformation
        \begin{eqnarray}
        &&\delta\bar\varphi=\frac{\partial\bar\varphi}
        {\partial\varphi}\,\delta\varphi,\,\,\,\,
        \bar\psi=-\frac1a\frac{\partial\bar a}
        {\partial\varphi}\,\delta\varphi+\frac{a}{\bar a}
        \frac{\partial\bar a}{\partial a}\,\psi,    \label{5.17}\\
        &&\bar\Pi_{\delta\varphi}=\frac{\partial\varphi}
        {\partial\bar\varphi}\,\Pi_{\delta\varphi}-
        \frac1a\frac{\partial a}
        {\partial\bar\varphi}\,\Pi_\psi,\,\,\,\,
        \bar\Pi_\psi=\frac{\bar a}{a}
        \frac{\partial a}{\partial\bar a}\,\Pi_\psi.  \label{5.18}
        \end{eqnarray}
The physical momentum $\bar p_{\delta\varphi}$ expresses in terms of phase
space momenta by the barred version of eq. (\ref{4.24}). Then, in view of
the above relations, one easily finds
        \begin{eqnarray}
        \bar p_{\delta\varphi}=\frac{\partial\varphi}
        {\partial\bar\varphi}\,p_{\delta\varphi}.       \label{5.20}
        \end{eqnarray}
This equation holds exactly for an arbitrary choice of the gauge condition
function $\chi(a,\varphi)$, and this is a corollary of the triangular form of
the transformation (\ref{5.16}). In our nonminimal model with $|\xi|\gg1$
this implies the following simple relation between the physical sectors in
two frames
        \begin{eqnarray}
        \delta\bar\varphi\simeq\sqrt{\frac3{4\pi}}m_P
        \frac{\delta\varphi}{\varphi},\,\,\,
        \bar p_{\delta\varphi}\simeq
        \sqrt{\frac{4\pi}3}\frac\varphi{m_P}\,
        p_{\delta\varphi}.                              \label{5.22}
        \end{eqnarray}

\subsection{Quadratic order currents}
\hspace{\parindent}
In the minisuperspace sector of the non-minimal model
$Q=(N(t),a(t),\varphi(t))$ the functional derivatives of the classical action
read
        \begin{eqnarray}
        &&\frac1{a^3\sqrt\gamma}\,\frac{\delta S}
        {\delta N}=6U(\varphi)\left(\frac1{a^2}+\frac{\dot a^2}{a^2}\right)
        +6U_\varphi(\varphi)\dot\varphi\frac{\dot a}a
        -\frac{\dot\varphi^2}2
        -V(\varphi),                                  \label{5.23}\\
        &&\frac 1{Na^3\sqrt\gamma}\,\frac{\delta S}
        {\delta\varphi}=
        -\ddot\varphi-3\frac{\dot a}a\dot\varphi
        +6U_\varphi(\varphi)\left(\frac1{a^2}
        +\frac{\ddot a}a
        +\frac{\dot a^2}{a^2}\right)-V_\varphi(\varphi),    \label{5.24}
        \end{eqnarray}
where dots are used to denote the parametrization invariant derivative
$\dot a\equiv da/Ndt,\,\dot\varphi\equiv d\varphi/Ndt$. Now we use the
perturbed ansatz (\ref{5.10}) for total minisuperspace variables in these
equations, $N^2\to N^2_{\rm tot}=a^2(1+2A)$, $a^2\to a^2_{\rm tot}
=a^2(1-2\psi)$, $\varphi\to\varphi_{\rm tot}=\varphi+\delta\varphi$,
and carefully expand the first order variations of the classical action
up to the second order in perturbations $(A,\psi,\delta\varphi)$ on
the {\em classical} background. The result reads as follows
        \begin{eqnarray}
        &&\frac 1{a\sqrt\gamma}\,\left[\,\frac{\delta S}
        {\delta N}\,\right]_2
        = A^2 \left(24U{\cal H}^2
        + 24U_{\varphi}{\cal H}\varphi'-2\varphi'^2\right) \nonumber\\
        &&\qquad\qquad\qquad\quad
        +A\psi\left(36U{\cal H}^2+36U_{\varphi}{\cal H}\varphi'
        -3\varphi'^2\right)
        +A\psi'\left(24U{\cal H}
        +12U_{\varphi}\varphi'\right)                     \nonumber\\
        &&\qquad\qquad\qquad\quad+A\delta\varphi
        \left(-12U_{\varphi}{\cal H}^2
        -12U_{\varphi\varphi}{\cal H}\varphi'\right)
         +A\delta\varphi'\left(-12U_{\varphi}{\cal H}
        +2\varphi'\right)                                 \nonumber\\
        &&\qquad\qquad\qquad\quad-12U\psi^2
        +\psi\psi'\left(12U{\cal H}
        +6U_{\varphi}\varphi'\right)+6U\psi'^2             \nonumber\\
        &&\qquad\qquad\qquad\quad+\psi\delta\varphi
        \left[-6U_{\varphi}(1+3{\cal H}^2)
        -18U_{\varphi\varphi}\varphi'
        +3a^2V_{\varphi}\right]                         \nonumber\\
        &&\qquad\qquad\qquad\quad+\psi'\delta\varphi
        \left(-12U_{\varphi}{\cal H}
        -6U_{\varphi\varphi}\varphi'\right)
        +\psi\delta\varphi'\left(-18U_{\varphi}{\cal H}
        +3\varphi'\right)
        -6U_{\varphi}\psi'\delta\varphi'                   \nonumber\\
        &&\qquad\qquad\qquad\quad+\delta\varphi^2
        \Big[3U_{\varphi\varphi}\Big(1+{\cal H}^2\Big)
        -\frac12 a^2V_{\varphi\varphi}
        +3U_{\varphi\varphi\varphi}{\cal H}\varphi'\Big]     \nonumber\\
        &&\qquad\qquad\qquad\quad
        +6U_{\varphi\varphi}{\cal H}
        \delta\varphi\delta\varphi'
        -\frac12\delta\varphi'^2,                          \label{5.30}\\
        &&\frac 1{a^2\sqrt\gamma}\,
        \left[\,\frac{\delta S}
        {\delta \varphi}\,\right]_2 =
        A^2\left(2a^2V_\varphi-12U_{\varphi}\right)
        +AA'\left(18U_{\varphi}{\cal H}-3\varphi'\right)
        \nonumber\\
        &&\qquad\qquad\qquad\quad+A\psi
        \left(6a^2V_\varphi-24U_{\varphi}\right)
        +(A\psi)'\left(18U_{\varphi}{\cal H}-3\varphi'\right)
        +6U_{\varphi}(A\psi')'
                 \nonumber\\
        &&\qquad\qquad\qquad\quad+A\delta\varphi
        \left[6U_{\varphi\varphi}(1-{\cal H}^2-{\cal H}')-
        a^2V_{\varphi\varphi}\right]
        +2{\cal H}A\delta\varphi'
                 \nonumber\\
        &&\qquad\qquad\qquad\quad
        -6U_{\varphi\varphi}{\cal H}A'\delta\varphi
        +(A\delta\varphi')'
                 \nonumber\\
        &&\qquad\qquad\qquad\quad-12U_{\varphi}\psi^2
        +\psi\psi'\left(18U_{\varphi}{\cal H}
        -3\varphi'\right)+6U_{\varphi}\psi\psi''
                \nonumber\\
        &&\qquad\qquad\qquad\quad+\psi\delta\varphi
        \left[-6U_{\varphi\varphi}(1+3{\cal H}^2+3{\cal H}')
        +3a^2V_{\varphi\varphi}\right]
        -18U_{\varphi\varphi}{\cal H}\psi'\delta\varphi
                 \nonumber\\
        &&\qquad\qquad\qquad\quad+6{\cal H}\psi\delta\varphi'
        +3(\psi\delta\varphi')'
        -6U_{\varphi\varphi}\psi''\delta\varphi
                 \nonumber\\
        &&\qquad\qquad\qquad\quad+\delta\varphi^2
        \left[3U_{\varphi\varphi\varphi}
        \Big(1+{\cal H}^2+{\cal H}'\Big)
        -\frac12 a^2V_{\varphi\varphi\varphi}\right].    \label{5.31}
        \end{eqnarray}
Here the unlabelled variables correspond to the minisuperspace background
in the conformal time gauge, $N=a,\,\,t=\eta$, primes denote derivatives
with respect to $\eta$ and ${\cal H}=a'/a$ denotes the conformal time
background Hubble constant. The background satisfies classical equations of
motion which were used to simplify the coefficients of the above quadratic
forms in $\Delta Q=(A,\psi,\delta\varphi)$.

\subsection{Quantum rolling force: effective action and minisuperspace
contributions}
\hspace{\parindent}
In the presence of the spatial densities of one-loop radiation currents
        \begin{eqnarray}
        &&j_N \equiv\frac 1{a^3\sqrt\gamma}
        \left<\,\left[\frac{\delta S}{\delta N}\,
        \right]_2\right>,                             \label{5.32}\\
        &&j_\varphi\equiv\frac 1{Na^3\sqrt\gamma}
        \left<\,\left[\frac{\delta S}{\delta \varphi}\,
        \right]_2\right>,                             \label{5.33}
        \end{eqnarray}
$N$ and $\varphi$ components of the effective equations of motion in
the non-minimal model read
        \begin{eqnarray}
        &&6U\left(\frac1{a^2}+\frac{\dot a^2}{a^2}\right)
        +6U_\varphi\dot\varphi\frac{\dot a}a
        -\frac{\dot\varphi^2}2
        -V+j_N=0,                                      \label{5.34}\\
        &&-\ddot\varphi-3\frac{\dot a}a\dot\varphi
        +6U_\varphi\left(\frac1{a^2}
        +\frac{\ddot a}a+\frac{\dot a^2}{a^2}\right)
        -V_\varphi+j_\varphi=0.                         \label{5.35}
        \end{eqnarray}

In view of eqs.(\ref{g1}) and (\ref{g10}) their $a$-component expresses in
terms of the above two ones, so that eqs.(\ref{5.34})-(\ref{5.35})
in a consistent manner exhaust the quantum dynamics of the mean fields.
Differentiating the first of them with respect to time one obtains the
system of two equations for $\ddot a$ and $\ddot\varphi$. Substituting the
solution of this system for $\ddot a$ into the second equation one finally has
the equation for the mean inflaton field with quantum contributions
to the friction term and the rolling force
        \begin{eqnarray}
        &&\ddot\varphi+\Big(3\,\frac{\dot a}a
        -\frac{a}{2\dot a}\,
        U_\varphi\,j_\varphi\Big)\dot\varphi
        -F(\varphi,a,\dot\varphi)=0,                    \label{5.36}\\
        &&F(\varphi,a,\dot\varphi)=
        \frac{2VU_\varphi-UV_\varphi}
        {U+3U_\varphi^2}-\frac12\dot\varphi^2\frac d{d\varphi}
        \ln(U+3U_\varphi^2)
        +F_{\rm loop}(\varphi,a,\dot\varphi,\dot a),   \label{5.36a}\\
        &&F_{\rm loop}(\varphi,a,\dot\varphi,\dot a)
        =\frac1{U+3U_\varphi^2}\left(Uj_\varphi
        -2U_\varphi\,j_N
        -\frac a{2\dot a}\frac{d\,j_N}{dt}\right).    \label{5.37}
        \end{eqnarray}

The first two terms in eq.(\ref{5.36a}) represent the classical rolling
force, the $\dot\varphi^2$ contribution belonging to the subleading
order of the slow roll expansion. As regards the quantum part, its radiation
currents in the one-loop approximation split into the contributions of the
quantum mechanical sector and the field sector of spatially inhomogeneous
modes, $j_{\rm 1-loop}=j^q+j^f$ (cf. eq.(\ref{2.12})). According to the
discussion of Sect. 3.1, see eqs.(\ref{2.24a})-(\ref{2.24c}), the $f$-part
of the current can be absorbed by
the replacement of the original classical action with the effective one
(\ref{2.24c}). This implies the replacement of the classical coefficient
functions $V(\varphi),U(\varphi),G(\varphi)$, (\ref{5.6})-(\ref{5.6b}),
by their effective counterparts
        \begin{eqnarray}
        S_{\rm eff}[g_{\mu\nu},\varphi]=
        \int d^4x\, g^{1/2}\left\{-V^{\rm eff}(\varphi)
        +U^{\rm eff}(\varphi)R
        -\frac12 G^{\rm eff}(\varphi)
        (\nabla\varphi)^2+...\right\},    \label{5.37a}
        \end{eqnarray}
and truncation of the (generally infinite) series to the first three terms.
This truncation is based on two assumptions -- the smallness of inflaton
derivatives due to the slow roll regime and smallness of $R/m^2_{\rm part}$
-- the curvature to particle mass squared ratio\footnote{The
Schwinger-DeWitt expansion involves the inverse powers of masses of
particles of constituent quantum fields. The latter acquire their
masses via the Higgs effect due to the interaction with the inflaton, so
that this ratio becomes order of magnitude $\lambda/|\xi|\ll 1$ \cite{efeq}.}.
Thus, with this approximation, the effective equations of motion in our
non-minimal model take the form of (\ref{5.34}) and (\ref{5.36}) with
$V_{\rm eff}(\varphi),U_{\rm eff}(\varphi),G_{\rm eff}(\varphi)$ replacing
$V(\varphi),U(\varphi),G(\varphi)$ and the radiation currents $j_N,j_\varphi$
saturated by the contribution of the quantum mechanical mode,
$j_N^q,j_\varphi^q$. The resulting rolling force in the leading order
of the slow roll expansion becomes the sum of the force induced by the
effective action, $F^{\rm eff}$, and the quantum mechanical force, $F^q$,
        \begin{eqnarray}
        &&F=F^{\rm eff}+F^q,   \\
        &&F^{\rm eff}=\frac{2V^{\rm eff}U_\varphi^{\rm eff}
        -U^{\rm eff}V_\varphi^{\rm eff}}
        {G^{\rm eff}U^{\rm eff}+3(U^{\rm eff}_\varphi)^2},   \label{5.37c}\\
        &&F^q=\frac1{U+3U_\varphi^2}\left(Uj^q_\varphi
        -2U_\varphi\,j_N^q
        -\frac a{2\dot a}\frac{d\,j_N^q}{dt}\right).    \label{5.37b}
        \end{eqnarray}

The one-loop calculation of $V_{\rm eff}(\varphi),U_{\rm eff}(\varphi)$ and
$G^{\rm eff}(\varphi)$ and the effect of $F^{\rm eff}$ on the inflationary
dynamics have been studied in \cite{efeq}. This effect is qualitatively
different for the no-boundary and tunneling cases and briefly looks as
follows. For the no-boundary state the one-loop corrections in the
distribution function add up to form the {\em full} Euclidean effective
action
        \begin{eqnarray}
        &&\rho_{NB}(\varphi)={\rm const}\exp[-
        \mbox{\boldmath$\SGamma$}(\varphi)],     \label{5.37d}\\
        &&{\mbox{\boldmath $\Gamma$}}(\varphi)=
        \mbox{\boldmath$I$}(\varphi)+
        {\mbox{\boldmath $\Gamma$}}^{\rm 1-loop}(\varphi)=
        -\frac{96\pi^2\,[\,U_{\rm eff}(\varphi)\,]^2}
        {V_{\rm eff}(\varphi)}
        +O(\hbar^2).                                \label{5.37e}
        \end{eqnarray}
Its value on the DeSitter instanton follows from that of the classical
Euclidean action, $\mbox{\boldmath$I$}(\varphi)=-96\pi^2 U^2/V$, by replacing
the classical coefficient functions $V(\varphi),U(\varphi)$ and $G(\varphi)$
by the effective ones, coinciding with those of the Lorentzian effective
action (\ref{5.37a}). Therefore, by the direct inspection of (\ref{5.37c})
one observes that the effective rolling force in the {\em no-boundary} case
is proportional to the derivative of the distribution
function\footnote{The explanation of this observation is simple. In the
minimal frame the rolling force is given by the gradient of the potential,
while the logarithm of the distribution function is inverse proportional to
it, the combination $V^{\rm eff}(\varphi)/[\,U^{\rm eff}(\varphi)\,]^2$
representing the minimal frame potential in terms of the non-minimal
objects (\ref{5.5}).}
        \begin{eqnarray}
        &&F^{\rm eff}_{NB}=\frac1{96\pi^2U^{\rm eff}}
        \frac{(V^{\rm eff})^2}
        {G^{\rm eff}U^{\rm eff}+3(U^{\rm eff}_\varphi)^2}
        \frac d{d\varphi}\ln\,\rho_{NB}(\varphi)      \nonumber\\
        &&\qquad\qquad\qquad=-\frac{\lambda m_P^2
        (1+\delta)}{48\pi\xi^2}\,\varphi
        \left(1-\frac{\varphi^2}{\varphi^2_I}\right)
        +O(1/|\xi|^3),                               \label{5.37f}
        \end{eqnarray}
and, thus, vanishes at the probability peak $\varphi_I$. The no-boundary
peak is realized for $1+\delta<0$, therefore the point $\varphi_I$ turns
out to be an attractor -- quantum terms in effective rolling force lock
the inflaton at its constant initial value and give rise to infinitely long
inflationary scenario with exactly DeSitter spacetime.

In the tunneling case, the distribution function is not related to
the overall effective action, because its tree-level part has a wrong sign.
The probability peak exists in the opposite range of the parameter
(\ref{delta}), $\delta>-1$, and the rolling force
        \begin{eqnarray}
        F^{\rm eff}_{T}=-\frac{\lambda m_P^2
        (1+\delta)}{48\pi\xi^2}\,\varphi
        \left(1+\frac{\varphi^2}{\varphi^2_I}\right)
        +O(1/|\xi|^3)                                \label{5.37g}
        \end{eqnarray}
has the quantum term which initially doubles the negative classical part.
Therefore, the inflaton starts slowly decreasing under the influence of
this force, and the tunneling state generates a finite inflation stage
with the estimated e-folding number (\ref{1.14}). These conclusions disregard
the contribution of the quantum mechanical radiation currents, and we proceed
to their calculation.

\subsection{Quantum state of the minisuperspace perturbations and their correlators}
\hspace{\parindent}
The calculation of quantum averages in the quadratic currents (\ref{5.30})
and (\ref{5.31}) requires the set of quantum correlators (\ref{2.25a}) of
bilinear combinations of minisuperspace disturbances
$\Delta\hat Q=(A,\psi,\delta\varphi)$ and their
derivatives. For this purpose we, first, need the reduced density
matrix of the inflaton field $\rho(\varphi,\varphi')$ in the non-minimal
model and, second, the expressions for the Heisenberg operators
$\Delta\hat Q(\eta)$ in terms of the Schrodinger operators of initial
perturbations and their momenta, $\delta\hat\varphi=\delta\varphi$,
$\hat p_{\delta\varphi}=\partial/i\partial(\delta\varphi)$.

As we know, the diagonal element of the density matrix has a quasi-gaussian
behavior (\ref{1.11}), which is, however, insufficient for averaging the
operators involving momenta. The necessary off-diagonal elements with
one-loop contributions of various
massive and massless fields have been calculated in \cite{decoh,fermion}.
It was shown that in the model with a big $|\xi|$ the initial density
matrix describes practically pure quantum state and
expresses in terms of the distribution function
        \begin{eqnarray}
        \rho(\varphi,\varphi')\,\Big|_{t=0}\simeq
        \sqrt{\rho(\varphi)}\sqrt{\rho(\varphi')},\,\,\,
        |\xi|\gg 1.                              \label{5.38}
        \end{eqnarray}
The explanation of this property \cite{decoh} is based on the fact that the
decoherence factor $D(\varphi,\varphi')$ by which the initial density matrix
differs from (\ref{5.38}) is a function of the arguments $m/H(\varphi)$ and
$m/H(\varphi')$ for a quantum field of a mass $m$. For large $|\xi|$ the
masses of particles generated by the Higgs effect give rise to big and
predominantly $\varphi$-independent ratio $m/H(\varphi)\sim\sqrt{|\xi|}$,
so that $D(\varphi,\varphi')\sim 1$. For massless fields a similar conclusion
can be drawn because for them the role of mass is played by the Hubble
constant $H(\varphi)$ of the quasi-DeSitter background.

In view of (\ref{1.11}), the effectively pure quantum state,
        \begin{eqnarray}
        \mbox{\boldmath$\Psi$}_{NB,T}(\varphi)\simeq
        \sqrt{\rho_{NB,T}(\varphi)},             \label{5.39}
        \end{eqnarray}
in the vicinity of the probability maximum, which is located at $\varphi_I$,
can, thus, be approximated by the gaussian packet of small quantum width
$\Delta$
        \begin{eqnarray}
        \mbox{\boldmath$\Psi$}_{NB,T}(\delta\varphi)
        \equiv\mbox{\boldmath$\Psi$}_{NB,T}(\varphi_I+\delta\varphi)
        =\frac1{(2\pi)^{1/4}\sqrt\Delta}
        \exp\left[-\frac{\delta\varphi^2}{4\Delta^2}\right].  \label{5.40}
        \end{eqnarray}

The operators of quantum disturbances in the $\delta\varphi$-representation,
acting on the wavefunction of the above type can be found by collecting
together several sets of equations derived above. First, we use the
equations (\ref{5.11})-(\ref{5.13}), relating $\Delta Q$ to
the Einstein frame perturbations $\Delta\bar Q$. Then, we apply the
barred version of equations (\ref{4.21})-(\ref{4.22}) to express the
minisuperspace perturbations $\Delta\bar Q$ in the Newton gauge as functions
of the invariants $\mbox{\boldmath$q$}$ and $\mbox{\boldmath$q$}'$ in the
minimal frame. Finally, we use the
set of equations (\ref{4.31a}), (\ref{4.32}) and (\ref{4.36})-(\ref{4.37})
with barred (minimal frame) potential and physical variables
to express these invariants in terms of $\delta\bar\varphi$ and
$\bar p_{\delta\varphi}$. The final result looks as follows
        \begin{eqnarray}
        &&A=-\frac1{3a\varphi'}\sqrt\frac{\kappa}{|\xi|}\left[
        \Big(\mbox{\boldmath$Q$}'_+ -\frac{3\varphi'}{\varphi}
        \mbox{\boldmath$Q$}_+\Big)\,\hat c_+
        +\Big(\mbox{\boldmath$Q$}'_- -\frac{3\varphi'}{\varphi}
        \mbox{\boldmath$Q$}_-\Big)\,
        \hat c_-\right],                                 \label{5.41}\\
        &&\psi=\frac1{3a\varphi'}\sqrt\frac{\kappa}{|\xi|}\left[
        \Big(\mbox{\boldmath$Q$}'_+ +\frac{3\varphi'}{\varphi}
        \mbox{\boldmath$Q$}_+\Big)\,\hat c_+
        +\Big(\mbox{\boldmath$Q$}'_-
        +\frac{3\varphi'}{\varphi}
        \mbox{\boldmath$Q$}_-\Big)\,\hat c_-\right],    \label{5.42}\\
        &&\delta\varphi=\frac1{\sqrt{6|\xi|}a\varphi'}
        \Big(\mbox{\boldmath$Q$}'_+ \hat c_+
        +\mbox{\boldmath$Q$}'_- \hat c_-\!\Big),         \label{5.43}
        \end{eqnarray}
where the operators $\hat c_{\pm}$ with the aid of (\ref{5.22})
read as
        \begin{eqnarray}
        &&\hat c_+=-\frac{1+\delta}{576\pi^3}\sqrt{\frac{2\pi}3}
        \frac{m_P^3}{|\xi|^2\varphi_I}
        \frac\partial{i\partial(\delta\varphi)},     \label{5.44}\\
        &&\hat c_-=\frac{1+\delta}{2\lambda\pi}
        \frac{m_P^3}{\varphi^3_I}
        \sqrt{\frac3{2\pi}}\delta\varphi.            \label{5.45}
        \end{eqnarray}

Now we are ready to find the quantum correlators necessary for the
radiation currents. We choose a symmetrized combination of disturbances
and their conformal time derivatives,
$\Delta Q_{1,2}=(\Delta\hat Q,\Delta\hat Q',\Delta\hat Q'')$, in the
definition of the correlator
        \begin{eqnarray}
        \big<\,\Delta Q_1\Delta Q_2\,\big>\equiv
        \frac12 \int d(\delta\varphi)
        \mbox{\boldmath$\Psi$}^*(\delta\varphi)
        (\Delta\hat Q_1\Delta\hat Q_2
        +\Delta\hat Q_2\Delta\hat Q_1)
        \mbox{\boldmath$\Psi$}(\delta\varphi),          \label{5.46}
        \end{eqnarray}
because in the Hermitian operators of quadratic currents (\ref{2.6}) the
products of operator valued disturbances automatically enter in symmetrized
form (in view of the symmetry of the 3-vertex function).
The further calculation of the correlators and radiation currents is
straightforward. However, the general answer that involves the basis
functions $\mbox{\boldmath$Q$}_\pm(\eta)$ for arbitrary $\eta$ is still
very complicated. Therefore, we separately consider the beginning of
the inflation epoch, $\eta=0$, and the late stationary stage of inflation.

\section{The onset of inflation}
\hspace{\parindent}
At the onset of inflation the basis functions $\mbox{\boldmath$Q$}_\pm(\eta)$
have a behaviour (\ref{4.34})-(\ref{4.35}). Using it in
eqs.(\ref{5.41})-(\ref{5.43}) with the operators $\hat c_\pm$ defined by
(\ref{5.44})-(\ref{5.45}) one easily obtains the initial equal-time
correlators with respect to the gaussian state (\ref{5.40}).

In the leading order of the slow roll expansion those correlators that do
not involve derivatives (potential type correlators) read
        \begin{eqnarray}
        &&\big<\,A^2\,\big>_0=-\big<\,A\psi\,\big>_0
        =\big<\,\psi^2\,\big>_0=
        \frac\lambda{288\pi^2|\xi|^2}\frac1f,          \nonumber\\
        &&\big<\,\psi\delta\varphi\,\big>_0=
        -\big<\,A\delta\varphi\,\big>_0=
        \frac{\lambda\varphi_I}{288\pi^2|\xi|^2}\frac1f,  \nonumber\\
        &&\big<\,\delta\varphi^2\,\big>_0=
        \frac{\lambda\varphi^2_I}{288\pi^2|\xi|^2}\frac1f,    \label{6.3}
        \end{eqnarray}
while the correlators of conformal time ``velocities'' (the kinetic
type correlators) equal
        \begin{eqnarray}
        &&\big<\,A'^2\,\big>_0=
        -\big<\,A'\psi'\,\big>_0=\big<\,\psi'^2\,\big>_0
        =\frac{\lambda}{288\pi^2|\xi|^2} f,            \nonumber\\
        &&\big<\,\psi'\delta\varphi'\,\big>_0=
        -\big<\,A'\delta\varphi'\,\big>_0=
        \frac{\lambda\varphi_I}{288\pi^2|\xi|^2} f,     \nonumber\\
        &&\big<\,\delta\varphi'^2\,\big>_0=
        \frac{\lambda\varphi^2_I}{288\pi^2|\xi|^2} f.    \label{6.4}
        \end{eqnarray}
As we see, these two groups of correlators differ by the power of a
special parameter $f$ which is inverse proportional to the square of quantum
dispersion of the inflaton field
        \begin{eqnarray}
        f\equiv\frac\lambda{48\pi^2|\xi|^2}
        \frac1{\kappa\Delta^2}
        =\left(\frac{\mbox{\boldmath $A$}}{16\pi^2}\right)^2
        \frac{|\xi|}{|1+\delta|}.                          \label{6.5}
        \end{eqnarray}
Such a dependence on $f$ reflects an obvious fact that the kinetic
type correlators (or correlators of momenta) in the gaussian state of the
form (\ref{5.40}) are inverse proportional to $\Delta^2$, and thus grow with
$\Delta\to 0$, while the potential type correlators (correlators of
coordinates) are proportional to $\Delta^2$, and thus decrease with the
narrowing of the gaussian peak.

The calculation of mixed correlators with one or two derivatives of the form
        \begin{eqnarray}
        \big<\,\Delta Q\Delta Q'\,\big>_0=
        \big<\,\Delta Q\Delta Q\,\big>_0 O(\epsilon^2),\,\,\,
        \big<\,\Delta Q\Delta Q''\,\big>_0=
        \big<\,\Delta Q\Delta Q\,\big>_0 O(\epsilon)     \label{6.6}
        \end{eqnarray}
shows that they belong to the subleading order in the slow roll parameter
(\ref{5.9}). Finally, the additional correlators with three derivatives,
which arise in the calculation of $dj^q_N/dt$ in the quantum rolling force
(\ref{5.37b}), express as
        \begin{eqnarray}
        &&\big<\,\psi''\delta\varphi'\,\big>_0=
        \big<\,\psi'\delta\varphi''\,\big>_0=-2{\cal H}
        \big<\,\psi'\delta\varphi'\,\big>_0,      \nonumber\\
        &&\big<\,\delta\varphi'\delta\varphi''\,\big>_0=
        -2{\cal H}\big<\,\delta\varphi'^2\,\big>_0,   \nonumber  \\
        &&\big<\,\psi'\psi''\,\big>_0=
        -2{\cal H}\big<\,\psi'^2\,\big>_0,           \label{6.7}
        \,\,\,\,\,\,\eta\to 0.
        \end{eqnarray}
Although they tend to zero in view of ${\cal H}(\eta)\to 0$ at $\eta\to 0$,
their contribution to the quantum rolling force is nontrivial because in
(\ref{5.37}) they are devided by $\dot a/a={\cal H}/a$.

Let us now go over to the calculation of radiation currents at $\eta=0$.
Within the slow roll approximation, $m_P^2/|\xi|\varphi^2\ll 1$,
$|\xi|\gg 1$, and in view of a particular form of classical coefficient
functions $V(\varphi), U(\varphi)$, the quadratic currents
(\ref{5.30})-(\ref{5.31}) are dominated by the following expressions
involving both the potential and kinetic terms
        \begin{eqnarray}
        &&j_N^q(0)=\frac{\lambda\varphi^4}4\left[-2\big<\psi^2\big>
        +\frac{10}\varphi\big<\psi\delta\varphi\big>
        -\frac{15}{\varphi^2}\big<\delta\varphi^2\big>
        +\big<\psi'^2\big>
        -\frac2\varphi\big<\psi'\delta\varphi'\big>\right],        \\
        &&j_\varphi^q(0)=\lambda\varphi^3\left[4\big<A\psi\big>
        +\frac7\varphi\big<\psi\delta\varphi\big>
        +\frac12\big<A'\psi'\big>\right].
        \end{eqnarray}
On using the tables of correlators above, the radiation currents,
contributing to the quantum rolling force, take the following final form
        \begin{eqnarray}
        &&j_N(0)=\frac{\lambda\varphi_I^4}4
        \frac{\lambda}{96\pi^2|\xi|^2}
        \left(\frac1f-\frac13 f\right),       \label{6.8} \\
        &&j_\varphi(0)=\lambda\varphi^3_I
        \frac{\lambda}{96\pi^2|\xi|^2}
        \left(\frac1f-\frac16 f\right),        \label{6.9}\\
        &&\frac a{2\dot a}
        \frac{d\,j_N}{dt}(0)=0.                \label{6.10}
        \end{eqnarray}

These quantities are strongly suppressed as compared to their classical
values, $-j_N=V(\varphi_I)\simeq\lambda\varphi_I^4/4$,
$-j_\varphi=V_\varphi(\varphi_I)\simeq\lambda\varphi_I^3$ by a very small

factor $\lambda/|\xi|^2\sim\Delta T^2/T^2\sim 10^{-10}$ related to the
CMBR anisotropy. Their sign crucially depends on the magnitude of the
parameter $f$, (\ref{6.5}), which in our model is likely to be very big,
$f\gg 1$. This follows from the estimate $N\geq 60$ on the e-folding number
(\ref{1.14}) and the value of $|\xi|\sim 10^4$ \cite{nonmin2}.
In this case, the terms proportional to $f\sim 1/\Delta^2$, generated by the
kinetic terms of the radiation currents, $\big<\Delta Q'\Delta Q'\big>$,
dominate and, in particular, result in
        \begin{eqnarray}
        \varepsilon^q(0)=-j_N(0)\simeq
        \frac{\lambda^2|1+\delta|}{18|\xi|^3}\,
        \frac{m_P^4}{(16\pi^2)^2}\ll m_P^4.
\label{6.11} \end{eqnarray} Interestingly, the sign of the quantum rolling
force due to the homogeneous mode is independent of the magnitude of $f$,
because in the leading order of the slow-roll expansion the contributions of
potential terms, $\Big<\Delta Q\Delta Q \Big>\sim 1/f$, completely cancel out
        \begin{eqnarray}
        F^q(0)=\frac{Uj^q_\varphi(0)-2U_\varphi j^q_N(0)}
        {U+3U_\varphi^2}
        \simeq \frac{\lambda\varphi^3_I}{36}
        \frac\lambda{96\pi^2|\xi|^3}f>0.                     \label{6.12}
        \end{eqnarray}
In view of the expressions for $f$ and $\varphi_I$, (\ref{1.9}), the
magnitude of this force is again much smaller than its classical counterpart
        \begin{eqnarray}
        F^q(0)=\frac{\lambda m_P^2}{48\pi|\xi|^3}\varphi_I
        \frac\lambda{144\pi^2}
        \frac{\mbox{\boldmath $A$}}{16\pi^2}
        \simeq |F^{\rm class}(0)| \frac\lambda{144\pi^2|\xi|}
        \frac{\mbox{\boldmath $A$}}{16\pi^2|1+\delta|}\ll
        |F^{\rm class}(0)|.                                \label{6.14}
        \end{eqnarray}
Therefore, for the tunneling state it gives a negligible contribution to the
effective force (\ref{5.37g}). For the no-boundary state, the initial
effective force (\ref{5.37f}) vanishes, but the only effect that the
positive $F^q(0)$ can produce in this case is that it shifts the equilibrium
point from $\varphi_I$ to slightly higher value of the inflaton $\varphi_*$,
$F^q(0)+F^{\rm eff}_{NB}(\varphi_*)=0$, at which again the system will
undergo endless inflation.

\section{Late stage of inflation}
\hspace{\parindent}
At late stationary stage of inflation the dynamics of the classical background
can be approximated by the ansatz
        \begin{eqnarray}
        &&a=\frac1{H(\varphi)}\cosh[H(\varphi)t],\,\,\,\,
        \varphi\simeq\varphi_I,    \label{7.3} \\
        &&H^2(\varphi)=
        \frac{V(\varphi)}{6U(\varphi)}
        \simeq\frac{\lambda \varphi^2}{12|\xi|}      \label{7.4}
        \end{eqnarray}
with the Hubble constant $H(\varphi)$ approximately linear in $\varphi$.
In the Einstein frame, it looks similar with the Hubble constant which is
practically independent of the inflaton $\bar H^2(\bar\varphi)
=8\pi\bar V(\bar\varphi)/3m_P^2\simeq\lambda m_P^2/96\pi|\xi|^2$,
the cosmic time parameters being related in both frames by
$\bar t\simeq t\sqrt{8\pi|\xi|}\varphi/m_P$.

The transition period between the onset of inflation and its steady stage
can be described by solving the inflaton equation with the approximately
constant rolling force $F$ and the friction term based on the ansatz
(\ref{7.3}) for $\dot a/a$
        \begin{eqnarray}
        &&\ddot\varphi+3H\tanh(Ht)\dot\varphi-F=0,  \nonumber\\
        &&\varphi(0)=\varphi_I,\,\,\,\dot\varphi(0)=0.   \label{7.5}
        \end{eqnarray}
For late times, $Ht\gg1$ (but not so late that the inflaton field evolves too
far from its initial value), the exact solution to this equation
        \begin{eqnarray}
        \varphi(t)=\varphi_I+\frac F{3H^2}\ln(\cosh Ht)
        +\frac F{3H^2}\tanh^2(Ht)                       \label{7.6}
        \end{eqnarray}
reads as an almost linear function of $t$
        \begin{eqnarray}
        \varphi(t)=\varphi_I+\frac{F}{3H}t
        +\frac F{3H^2}(1-\ln2)
        +O\left(e^{-2Ht}\right).                     \label{7.7}
        \end{eqnarray}
This behaviour corresponds to neglecting the $\ddot\varphi$ term in the
inflaton equation of motion and solving it for $\dot\varphi$,
$\dot\varphi\simeq F/3H$. In our model with the classical rolling force
(\ref{1.13}), $\dot\varphi\simeq -4\varphi H\epsilon/3\ll H\varphi$ with
$\epsilon\sim m_P^2/|\xi|\varphi^2$ -- the slow roll smallness parameter
(\ref{5.9}). Thus, in the lowest order of the slow roll approximation the
inflaton field remains constant.

Let us study the behaviour of the basis functions
$\mbox{\boldmath$Q$}_\pm(\eta)$ for $Ht\gg1$. To begin with, note that
for late times corresponding to exponentially large scale factor the
potential terms in the wave equation for $\mbox{\boldmath$Q$}_\pm(\eta)$,
(\ref{4.33}), can be discarded. The first one,
$\kappa\bar\varphi'^2/2=O(\epsilon^2)$, is small in view of the slow roll
regime and the second one, the spatial curvature term $3$, is small compared
to the kinetic terms growing with $a$, $d^2/d\eta^2\sim a^2 d^2/dt^2$.
Therefore, at late times this equation simplifies to
        \begin{eqnarray}
        \left(-\frac{d^2}{d\eta^2}+2\frac{\bar\varphi''}{\bar\varphi'}
        \frac d{d\eta}\right)
        \mbox{\boldmath$Q$}_\pm(\eta)=0,        \label{7.8}
        \end{eqnarray}
and has two explicit solutions $\mbox{\boldmath$Q$}_\pm$
        \begin{eqnarray}
        &&\mbox{\boldmath$Q$}_-(\eta)=1,    \label{7.9a}\\
        &&\mbox{\boldmath$Q$}_+(\eta)=N_+
        \int_0^\eta d\tilde\eta\,
        \bar\varphi'^2(\tilde\eta),\,\,\,\,
        N_+\simeq\left(\frac{8\pi|\xi|\varphi^2_I}{m_P^2}\right)^2
        \frac\pi{m_P^2(1+\delta)^2},             \label{7.9}
        \end{eqnarray}
compatible with the expansions (\ref{4.34})-(\ref{4.35}) at early times
$\eta\to 0$. (The mismatch between the constant function (\ref{7.9a})
and eq.(\ref{4.35}) has a simple explanation: the term
$-3\eta^2/2$ in $\mbox{\boldmath$Q$}_-$ of eq.(\ref{4.35}) is induced
by the curvature term which we discard at late times, while the $O(\eta^3)$
corrections are due to the perturbation $\kappa\bar\varphi'^2/2$.) In terms
of the cosmic time in the original frame, $\mbox{\boldmath$Q$}_+$ represents
a well known growing mode \cite{Starobinsky} which for late times in the
slow roll approximation reads as
        \begin{eqnarray}
        \mbox{\boldmath$Q$}_+=\frac13\sinh Ht.       \label{7.10}
        \end{eqnarray}
Thus at late times the operator of the invariant cosmological perturbation is
dominated by the growing mode

        \begin{eqnarray}
        \hat{\mbox{\boldmath$Q$}}(t)=\hat c_+\mbox{\boldmath$Q$}_+(t)
        +\hat c_-\mbox{\boldmath$Q$}_-(t)\simeq
        \hat c_+\mbox{\boldmath$Q$}_+(t),\,\,\,\,Ht\gg 1,   \label{7.11}
        \end{eqnarray}
and in the Newton gauge all the minisuperspace perturbations express in terms
of one operator $\hat c_+$ defined by the momentum $\hat p_{\delta\varphi}$,
(\ref{5.44}),
        \begin{eqnarray}
        &&\bar A=\bar\psi=\frac{\kappa\mbox{\boldmath$Q$}}{2\bar a(t)}
        \simeq\sqrt{\frac{\pi\lambda}{54}}
        \frac{\hat c_+}{m_P|\xi|},                    \nonumber\\
        &&\delta\bar\varphi=
        \frac{\dot{\mbox{\boldmath$Q$}}}{\bar a(t)\dot{\bar\varphi}(t)}
        \simeq-\frac{\sqrt{2\lambda}}3
        \frac{\pi\varphi^2}{m_P^2(1+\delta)}\hat c_+.     \label{7.12}
        \end{eqnarray}
Since $\delta\bar\varphi$ contains $\dot{\bar\varphi}$ in the denominator,
all the other perturbations in the minimal frame are much smaller in
magnitude, $(\bar A,\,\bar\psi)\sim O(\epsilon)\delta\bar\varphi/m_P\ll
\delta\bar\varphi/m_P$. Therefore, in view of eqs.(\ref{5.11})-(\ref{5.13})
the perturbations in the non-minimal frame read
        \begin{eqnarray}
        A\simeq -\psi=\simeq\sqrt{\frac{4\pi}3}
        \frac{\delta\bar\varphi}{m_P},\,\,\,\delta\varphi\simeq
        \sqrt{\frac{4\pi}3}\frac{\varphi}{m_P}
        \delta\bar\varphi,\,\,\,Ht\gg 1.                \label{7.13}
        \end{eqnarray}

Another important property of the perturbations in both frames is that to
the leading order in slow roll they are constant in time for $Ht\gg1$.
This follows from eqs.(\ref{7.12}) containing the exponentially growing
functions of time in both of its numerator and denominator (respectively
$\mbox{\boldmath$Q$}$ and $\bar a$). As a result, the time derivatives
of perturbations belong to the subleading order of the slow roll expansion,
$\Delta\dot Q\equiv(\dot A,\dot\psi,\delta\dot\varphi)
=O(\epsilon)\Delta Q$.

Thus, we arrive at the following list of correlators at late times. The
potential type correlators read
        \begin{eqnarray}
        &&\big<\,A^2\,\big>=\big<\,\psi^2\,\big>
        =-\big<\,A\psi\,\big>=\frac{\lambda}{2592\pi^2|\xi|^2}f, \nonumber\\
        &&\big<\,\psi\delta\varphi\,\big>=
        -\big<\,A\delta\varphi\,\big>=
        \frac{\lambda\varphi}{2592\pi^2|\xi|^2}f,   \nonumber\\
        &&\big<\,\delta\varphi^2\,\big>=
        \frac{\lambda\varphi^2}{2592\pi^2|\xi|^2}f,          \label{7.14}
        \end{eqnarray}
where the parameter $f$ is given by eq.(\ref{6.5}), while the kinetic type
correlators are negligibly small
        \begin{eqnarray}
        &&\big<\,\Delta Q\Delta Q'\,\big>=
        O(\epsilon)\,aH\big<\,\Delta Q\Delta Q\,\big>,  \nonumber\\
        &&\big<\,\Delta Q'\Delta Q'\,\big>=
        O(\epsilon^2)\,(aH)^2\big<\,\Delta Q\Delta Q\,\big>,  \nonumber\\
        &&\big<\,\Delta Q\Delta Q''\,\big>=
        O(\epsilon)\,(aH)^2\big<\,\Delta Q\Delta Q\,\big>,  \nonumber\\
        &&\big<\,\Delta Q'\Delta Q''\,\big>=
        O(\epsilon^2)\,(aH)^3\big<\,\Delta Q\Delta Q\,\big>   \label{7.15}
        \end{eqnarray}

In view of these relations, the terms that give the leading contribution
to radiation currents are exhausted by a small fraction of terms in
the equations (\ref{5.30})-(\ref{5.31}). They include only the potential
type correlators and read
        \begin{eqnarray}
        &&j_N^q=\frac{\lambda\varphi^4}4\left[4\big<A^2\big>
        +6\big<A\psi\big>
        -\frac4\varphi\big<A\delta\varphi\big>+
        \frac6\varphi\big<\psi\delta\varphi\big>
        -\frac5{\varphi^2}
        \big<\delta\varphi^2\big>\right],            \label{7.16} \\
        &&j_\varphi^q=\lambda\varphi^3\left[2\big<A^2\big>
        +6\big<A\psi\big>
        -\frac8\varphi\big<A\delta\varphi\big>+
        \frac6\varphi\big<\psi\delta\varphi\big>
        -\frac3{\varphi^2}
        \big<\delta\varphi^2\big>\right].             \label{7.17}
        \end{eqnarray}
The resulting radiation currents, thus, equal
        \begin{eqnarray}
        &&j_N^q=\frac{\lambda\varphi^4}4
        \frac{\lambda}{864\pi^2|\xi|^2}f,               \label{7.18}\\
        &&j_\varphi^q=\frac{\lambda\varphi^3}2
        \frac{\lambda}{864\pi^2|\xi|^2}f, \,\,\,\,\,Ht\gg 1   \label{7.19}
        \end{eqnarray}
Similarly to the onset of inflation, eqs.(\ref{6.8})-(\ref{6.9}), they are
strongly suppressed relative to the classical values,
$-V(\varphi)\simeq-\lambda\varphi^4/4$ and
$-V_\varphi(\varphi)\simeq-\lambda\varphi^3$ by the factor
$\lambda/|\xi|^2\sim 10^{-10}$. In absolute units, the energy density of the
quantum mechanical mode is given by
        \begin{eqnarray}
        \varepsilon^q=-j^q_N\simeq
        -\frac{\lambda^2|1+\delta|}{54|\xi|^3}\,
        \frac{m_P^4}{(16\pi^2)^2},
        \,\,\,\,|\varepsilon^q|\ll m_P^4.                \label{7.20}
        \end{eqnarray}

Note that, in contrast to the onset of inflation, this energy is negative.
Apparently, this is a manifestation of the ghost nature of the invariant
physical mode $\mbox{\boldmath$q$}$, whose kinetic term enters the action
(\ref{4.16}) with the wrong sign. In the slow roll approximation with
negligible $\dot\varphi$ a constant energy density (\ref{7.20}) in view of
eq.(\ref{2.11a}) -- the conservation law for radiation current -- generates
the effective equation of state of the inflaton excitation mode
        \begin{eqnarray}
        \varepsilon^q+p^q=0,\,\,\,
        \varepsilon^q<0, \,\,\,Ht\gg 1,               \label{7.21}
        \end{eqnarray}
which in the absence of other sources would maintain the Anti-DeSitter
spacetime. Similar equation of state at the onset of inflation corresponds
to the DeSitter case, so that the inflaton quantum excitation undergoes
a sort of phase transition reversing the sign of its energy density.

In view of the relation between the radiation currents,
$j_N^q=\varphi j_\varphi/4+O(\epsilon)$, and approximately constant
value of $j_N^q$, $dj_N^q/dt=O(\epsilon)$, the quantum rolling force in
eq.(\ref{5.37}),
        \begin{eqnarray}
        F^q\simeq\frac1{6|\xi|\varphi}
        \left(\varphi j^q_\varphi
        -4j^q_N
        -\frac1{2H}\frac{d\,j^q_N}{dt}\right)=O(\epsilon),
        \end{eqnarray}
vanishes in the leading order of the slow roll expansion. Thus, similarly to
the onset of inflation, at late times of inflation epoch $F^q$ does not
qualitatively change the cosmological evolution.

\section{Conclusions}
\hspace{\parindent}
We have developed a general framework for effective equations of
inflationary dynamics in quantum cosmology and for their
quantum Cauchy problem with no-boundary and tunneling quantum states.
This framework combines the Euclidean effective action method and the method
of direct quantum averaging for calculations of two distinctly different
parts of radiation currents -- contributions of the
field theoretical and quantum mechanical (minisuperspace) sectors of the
system. We focus on the latter and show that its calculation is based on
explicit physical reduction for the spatially homogeneous cosmological
perturbation. Because of the ghost nature of this perturbation, its
effect is not related to the conventional analytic continuation from the
Euclidean spacetime. Rather, in the model of strongly coupled non-minimal
inflaton it originates from the quasi-gaussian state which incorporates
the tree-level and one-loop effects on the DeSitter instanton. It is, thus,
irrelevant to the DeSitter invariant Euclidean vacuum and cannot be obtained
by analytic continuation from the Euclidean section of spacetime. This
means that the universality of analytic continuation methods of \cite{HHT}
should not be overestimated -- they apply to spatially inhomogeneous,
particle like excitation but may fail for minisuperspace cosmological
modes.

Unfortunately, the dynamical contribution of the quantum mechanical mode to
effective equations turned out to be disappointingly small -- it is strongly
dominated by the effective rolling force (contributed on equal footing by
the classical term and the one-loop term due to the inhomogeneous modes).
The property of its strong suppression by powers of $1/|\xi|\ll 1$ was
actually conjectured in \cite{efeq}, and now it is quantitatively confirmed.
Thus, the inflaton mode cannot change the
dynamical predictions in spatially closed model with strong non-minimal
coupling. As a model of the low-energy quantum origin of the Universe
only the tunneling state remains observationally justified, because the
no-boundary wavefunction generates infinitely long inflationary stage.
The role of this mode should not, however, be underestimated, because its
effect is model dependent, and might be important in other models generating
initial conditions for inflation \cite{BN}. Moreover, the quantum inflaton
mode simulates the DeSitter and Anti-DeSitter effective equations of state,
$\varepsilon+p=0$, respectively at the onset of inflation and at late times.
The sign of its energy density contribution can change depending on the
balance of the potential and kinetic terms of this ghost mode. Therefore,
it is not quite clear at the moment, what can the role of this mode be at
post inflationary epoch. A natural question arises if this mode can be
responsible for the present day observable acceleration of the Universe
\cite{Lambda} as an alternative to quintessence \cite{quint} or be
capable of inducing DeSitter-anti-DeSitter phase transitions in cosmology?
This question is subject to further studies \cite{BN}.

\section*{Acknowledgements}
\hspace{\parindent}
This work was supported by the Russian Foundation for Basic Research under
the grant No 99-02-16122. The work of A.O.B. was also supported by the
grant of support of leading scientific schools No 00-15-96699. The work of
D.V.N. was supported by the grant of support of leading scientific schools
No 00-15-96566. This work has also been supported in part by the Russian
Research program ``Cosmomicrophysics''.

\end{document}